\documentclass[]{interact}
\usepackage[caption=false]{subfig}% Support for small, `sub' figures and tables

\usepackage[numbers,sort&compress]{natbib}% Citation support using natbib.sty
\bibpunct[, ]{[}{]}{,}{n}{,}{,}% Citation support using natbib.sty
\theoremstyle{plain}% Theorem-like structures provided by amsthm.sty

\theoremstyle{definition}

\theoremstyle{remark}

\usepackage{bookmark}
\usepackage{amsmath}
\usepackage{amssymb}
\usepackage{enumitem}
\usepackage{booktabs}%% for tables
\usepackage{xcolor, colortbl}
\usepackage{float}
\usepackage{relsize}%% for fonts mathrm
\usepackage{graphics}
\usepackage{longtable}
\usepackage{algorithm}
\usepackage{algpseudocode}
\usepackage{algorithmicx}

\usepackage[modulo]{lineno}

\setpagewiselinenumbers

\def\spacingset#1{\renewcommand{\baselinestretch}{#1}}
\makeatletter
\newcommand*{\trasp}{%
  {\mathpalette\@trasp{}}%
}
\newcommand*{\@trasp}[2]{%
  ^{\raisebox{\depth}{$\m@th#1\scriptstyle{\intercal}$}}%\mathsf{T}
}
\makeatother

\newcommand{\trasps}[1]{^{\raisebox{\depth}{$\scriptstyle{\intercal}$}}_{#1}} %use without overset symbol
\newcommand{\traspd}[1]{^{\raisebox{-1.0mm}{$\scriptstyle{\intercal}$}}_{#1}} %use with overset symbol
\newcommand{\matinv}{^{\raisebox{+1.0mm}{$\scriptstyle{{-1}}$}}} %use for matrices
\newcommand{\subd}{{[d]}}

\newcommand{\ba}{{\mathbf{a}}}
\newcommand{\bc}{{\mathbf{c}}}
\newcommand{\bp}{{\mathbf{p}}}

\newcommand{\bt}{{\mathbf{t}}}

\newcommand{\bv}{{\mathbf{v}}}
\newcommand{\bx}{{\mathbf{x}}}

\newcommand{\dwj}{{\overset{\boldsymbol{.}}{\mathbf{w}}_j}}

\newcommand{\daj}{{\overset{\boldsymbol{.}}{\mathbf{a}}_j}}
\newcommand{\dajT}{{{\overset{\boldsymbol{.}}{\mathbf{a}}\traspd{j}}}}

\newcommand{\dc}{{\overset{\boldsymbol{.}}{\mathbf{c}}}}
\newcommand{\dcj}{{\overset{\boldsymbol{.}}{\mathbf{c}}_j}}

\newcommand{\dQj}{{\overset{\boldsymbol{.}}{\mathbf{Q}}_j}}
\newcommand{\dQjT}{{{\overset{\boldsymbol{.}}{\mathbf{Q}}\traspd{j}}}}

\newcommand{\btp}{{\overset{\sim}{\mathbf{p}}}}

\newcommand{\btP}{{\overset{\sim}{\mathbf{P}}}}

\newcommand{\dvj}{{\overset{\boldsymbol{.}}{\mathbf{v}}_j}}

\newcommand{\btV}{{\overset{\sim}{\mathbf{V}}}}

\newcommand{\bA}{{\mathbf{A}}}
\newcommand{\bB}{{\mathbf{B}}}
\newcommand{\bC}{{\mathbf{C}}}

\newcommand{\bE}{{\mathbf{E}}}
\newcommand{\bI}{{\mathbf{I}}}
\newcommand{\bL}{{\mathbf{L}}}
\newcommand{\bQ}{{\mathbf{Q}}}

\newcommand{\bS}{{\mathbf{S}}}
\newcommand{\bO}{{\mathbf{O}}}
\newcommand{\bP}{{\mathbf{P}}}
\newcommand{\bT}{{\mathbf{T}}}
\newcommand{\bX}{{\mathbf{X}}}
\newcommand{\bXT}{{\mathbf{X}\trasp}}

\newcommand{\bV}{{\mathbf{V}}}
\newcommand{\bU}{{\mathbf{U}}}

\newcommand{\bl}{{\mathbf{l}}}

\newcommand{\bLa}{{\mathbf{\Lambda}}}

\newcommand{\brj}{{\mathbf{r}_j}}

\newcommand{\bajT}{{\mathbf{a}\trasps{j}}}

\newcommand{\bpj}{{\mathbf{p}_j}}
\newcommand{\bpjT}{{\mathbf{p}\trasps{j}}}

\newcommand{\hpj}{{\hat{\mathbf{p}}}_j}

\newcommand{\dX}{{\overset{\boldsymbol{.}}{\mathbf{X}}}}
\newcommand{\dXj}{{\overset{\boldsymbol{.}}{\mathbf{X}}_j}}
\newcommand{\dXjT}{{\overset{\boldsymbol{.}}{\mathbf{X}}\traspd{j}}}

\newcommand{\vexp}{\text{vexp}}

\DeclareMathOperator*{\argmin}{arg\,min}
\DeclareMathOperator*{\argmax}{arg\,max}

\newcommand{\Li}{\text{L}_1}
\newcommand{\Lt}{\text{L}_2}
\newcommand{\Lmax}{\text{L}_\infty}

\newcommand\bsln{1.5} %% see <--- controls the
\spacingset{\bsln}
\begin{document}
\articletype{RESEARCH ARTICLE}% Specify the article type or omit as appropriate

\title{SIMPCA: A framework for rotating and sparsifying principal components}% coefficients
%\maketitle
\author{
    \name{Giovanni Maria Merola\thanks{CONTACT G. M. Merola. Email: merolagio@gmail.com}}
    \affil{Department of Mathematical Sciences, Xi'an Jiaotong-Liverpool University, Suzhou, PRC\\
    Published in the Journal of Applied Statistics\\
    DOI: 10.1080/02664763.2019.1676404
}
}
 \maketitle

\begin{abstract}
We propose an algorithmic framework for computing sparse components from rotated principal components. This methodology, called SIMPCA, is useful to replace the unreliable practice of ignoring small coefficients of rotated components when interpreting them. The algorithm computes genuinely sparse components by projecting rotated principal components onto subsets of variables. The so simplified components are highly correlated with the corresponding components. By choosing different simplification strategies different sparse solutions can be obtained which can be used to compare alternative interpretations of the principal components. We give some examples of how effective simplified solutions can be achieved with SIMPCA using some publicly available data sets.
\end{abstract}

%\linenumbers
\section{Introduction}
Dimensionality reduction methods (DRMs) aim to identify a lower dimensional structure underlying a set of observed variables.
Given an $n\times p$  matrix of $n$ observations of $p$ mean-centred variables, $\bX$, linear DRMs aim to approximate it with a set of $d < p$ linear combinations of its columns, $\bT = \bX\bA$, where $\bA$  is a $p \times d$ matrix of \emph{coefficients}. The \emph{rank-d} linear DRM model is
\begin{equation}\label{eq:pcamod}
    \bX = \bT \bL\trasp + \bE,
\end{equation}
where, the $n\times p$ matrix $\bE$ is the random error matrix as usual; the columns of the matrix $\bT$, $\bt_j$, are called generically \emph{components} and the $d\times p$ matrix $\bL$ is traditionally called matrix of \emph{loadings}. The loadings show how each variable depends (\emph{loads}) on the components and they play different roles in different DRMs.
In recent literature, the term loadings is used also for the coefficients $\bA$, we prefer to keep the distinction in order to avoid ambiguities when referring to less recent works.

Principal component analysis, PCA, is one of the most frequently used DRM methods for the analysis and exploration of multivariate data. The unique PCA solutions, called \emph{principal components} (pcs) which we denote as $\bP = \bX\bV$, are obtained by minimising the Frobenious ($\text{L}_2$) norm of the residuals $\bE$ under the constraint that they are mutually orthogonal.
The same solutions can be obtained from a number of problems. An important equality is:
\begin{equation}\label{eq:pcaopt}
    \bV = \begin{cases}
            \argmin\limits_{\bA\in\Re^{p\times d}} ||\bX -  \bT \bL\trasp||^2 \\
            \text{Subject to}\ \bt\trasps{i}\bt_j = 0,\; i < j\leq d
        \end{cases}
    =
\begin{cases}
        \argmax\limits_{\ba_j\in\Re^{p}} \sum_{j=1}^d \frac{\bajT\bX\trasp\bX\ba_j}{\bajT\ba_j},\, j =1,\ldots,d\\
% j = 1, \ldots, d\\
        \text{Subject to}\ \bajT\ba_i = 0,\;  i< j\leq d
    \end{cases}
\end{equation}
where a bold lower case symbol $\bc_j$ denotes the $j$-th column of a matrix $\bC$.
In PCA the matrix of loadings is completely determined by the matrix of coefficients and it is equal to $\bL\trasp = (\bP\trasp\bP)\matinv\bP\trasp\bX = \bV\trasp$.
The formal optimisation is carried out for all $1\leq d <p$ and the resulting components reproduce, or \emph{explain}, sequentially as much as possible variance of the data. The \emph{variance explained} by each pc is defined as $\vexp(\bpj) = ||\bp_j(\bpjT\bpj)^{-1}\bpjT\bX||^2$.
The least squares (LS) minimisation on the left hand side \cite[Pearson,][]{pea} is  equivalent to the maximisation of the variance of the components on the right hand side \cite[Hotelling,][]{hot} only for the optimal PCA solution. When different constraints are imposed on the components, the
two formulation are not equivalent. It should be noted that the PCA solutions are invariant to changes of scale of the coefficients, while there exists a widespread misconception about the pcs coefficients that they should always be scaled to unit $\Lt$ norm. Throughout this paper the term \emph{coefficients} refers to  values scaled to unit $\Lt$ norm.

%%%%%%%%%%%%%%%%%%%%%%%%%%%%%  USAGE OF PCA ===========================================
%\centerline{========== USAGE OF PCA =============}
\subsection{Usage of PCA}
The pcs can be used for different purposes. In some applications they are used for outliers or pattern detection (usually by visually inspecting the first few pcs), data compression, data whitening, etc. This usage is justified by the property of the pcs of retaining the most possible variance of the data.

In other cases, the pcs are used as estimates of nonobservable \emph{latent} variables useful to explain vaguely defined concepts. For example, the pcs could be used to estimate an index of ``intelligence'' from a battery of IQ tests, or an index of ``welfare'' from household survey data, or an index of ``quality'' from a set of products' measurements.
% Such indices can then be also used to classify the observed subjects into classes of latent categorical variables, such as quality grade of computer chipsets, or households eligible for income support, face and signature recognition, for example.

PCA is not a predictive method because its objective is to reproduce as much variance of the data as possible, with no reference to external variables. However, in some cases the estimate of a latent variable is validated against an external variable which is believed to be correlated with it. In this case, the ordering and optimality of the pcs are not necessarily relevant and components defined by combinations of some of the pcs may be better for this purpose.
When the PCA is used to estimate latent components, the analysis includes the interpretation of the pcs, that is, the identification of the key variables that define the latent constructs. Identifying key variables is also important when the components are used for diagnostic purposes, for process malfunctioning \cite{kou}, for example.

%\centerline{============ SIMPLICITY =========}
\subsection{Simplicity}
Since the pcs are combinations of many of the original variables they are difficult to interpret and different methods are used to simplify them. Different definitions of ``simplicity'' have been proposed. The most commonly used one is that of \emph{sparseness}, for which more useful solutions would be \emph{sparse}, that is a combinations of few of the original variables. Sparse methods have recently received great attention in the statistical literature, also thanks to the ``bet on sparseness'' principle \cite{has}. On one hand, sparseness is intuitively appealing and can easily be expressed in mathematical terms by constraining the \emph{cardinality} of the coefficients (the number of nonzero coefficients) to be low. On the other hand, interpretability is a vague concept, which depends on subjective judgement, Therefore, some sparse components could be judged to be not meaningful while others are found to be easier to interpret. As an example, consider two sparse components computed from a set of 16 baseball players' statistics (see Section \ref{sec:baseball} for details):
\begin{description}
  \item[Component a]
  \begin{align*}
  &14\%(\text{years in the major leagues}) + 17\%(\text{times at bat in career})\ + \\
  &13\%(\text{hits in career}) + 11\%(\text{home runs in career}) + 13\%(\text{runs in career})\ +\\
   &15\%(\text{runs batted-in in career}) + 17\%(\text{walks in career})
  \end{align*}
  \item[Component b]
  \begin{align*}
    &17\%(\text{runs in 1986}) + 15\%(\text{runs batted-in in 1986})\ + \\
    &25\%(\text{times at bat in career}) + 26\%(\text{runs batted-in in career})\ +\\
     &17\%(\text{walks in career}).
  \end{align*}
\end{description}
Component ``a'' is a combination of seven (all available) career statistics and component ``b'' is a combination of two season 1986 statistics and three career statistics. Some analysts may consider component ``a'' to be the more interpretable one, because it summarises a player's career performance. Others may consider component ``b'' to be more meaningful because it gives a comprehensive summary of the performance of a player using few key statistics. This means that, interpretable components cannot be found by simply applying a sparsifying method but that analysts prefer to have a set of simplified solutions from which to chose the ones that best lend themselves to their interpretation.

% The number of nonzero coefficients is called \emph{cardinality} of the sparse coefficients.

The first and simplest approach used to simplify the interpretation of the pcs coefficients is to \emph{threshold} them, that is ignore those which have absolute value lower than a given threshold. The practice of thresholding the coefficients can give misleading results \cite{cad}. In fact, large coefficients usually correspond to highly correlated variables and their values would be different if the smaller coefficients were truly zero. In spite of this, thresholding is commonly used to enhance the interpretability of the pcs.

%\centerline{======== ROTATION ===========}
\subsection{Rotation of components}
Rotating the components was proposed as a means to improve the interpretability of the loadings. A great many different algorithms for computing rotations have been proposed \cite[see][for an accessible review]{bro}. These algorithms intend to increase the separation between ``large'' and ``small'' loadings so as to facilitate their interpretation by tresholding. Rotation of the pcs was also suggested  for simplifying their interpretation \cite[among others,][]{ric, hen, jol, jac, kie}.
However, this procedure is not unanimously accepted because rotation destroys the sequential optimality of the pcs and, in some cases, their orthogonality. A discussion of the effect of rotating pcs coefficients under different scalings can be found in \cite{jol95}.

The idea of rotating the components derives from the indeterminacy of Model \ref{eq:pcamod}, by which the reduced rank decomposition of the data matrix  is invariant to postmultiplication of the coefficients by an invertible matrix. That is, if $\bO$ is a $d\times d$ invertible matrix\footnote{$\bO$ does not need to be a rotation matrix in a strict sense (i.e. does not need to be an element of the orthogonal group $SO(d)$, defined in  \cite{bac}, for example). In this context, rotation indicates a change of basis subsequent to that operated by the DRM.}, Model \ref{eq:pcamod} can equivalently be written as
\begin{equation}\label{eq:pcamodrot}
    \bX = \bT \bL\trasp + \bE = \bT\bO\bO\matinv \bL\trasp +  \bE,
\end{equation}
where $(\bT\bO)$ are the rotated components. The rotation is called \emph{orthogonal} if the matrix $\bO$ is such that $\bO\trasp\bO$ is diagonal and \emph{oblique} otherwise.
%Maybe Say that when rotating the pcs we should require that $\bO\trasp\bT\trasp\bT\bO$ is diagonal.

%\centerline{========= ROTATION FOR PREDICTION =========}

%\centerline{============ PCs as PREDICTORS =========}
In some cases the pcs are used to predict variables external to the dataset, for example to compute biased linear regression estimates when the observed explanatory variables are highly correlated. In other cases, as mentioned above, the pcs are used to estimate nonobservable latent variables which are validated with respect to variables external to the dataset. Since the optimality of the pcs is not relevant to external variables, rotated pcs can be more useful in a predictive context, \citep[e.g.][]{jaco, die}. Naturally, since also rotation is carried out without reference to any external variable, some rotation procedures yield better predictors than others. Deciding which rotation criterion produces more useful components requires to compare different solutions and subjective judgement. In some cases rotation may not improve the prediction of a specific variable or the interpretability of the component.

%\centerline{======== SPCA ================}
% SPCA methods optimise different functions which "break out" from the inflexibility of the pcs Least Square optimality. However, they, in turn, are not flexible, in the sense that they optimise the SPCA problem under the chosen penalty and the user has little room for tuning them. Such an approach can be very useful in highly dimensional problems, when visual inspection of the results is difficult. Also, their implementation requires sophisticated programming.
%
% The first method used to simplify pcs was to rotate them. This consists of multiplying a set of pcs by an invertible matrix and then threshold the loadings. Rotation in most cases increases the distance between \emph{small} and \emph{large} loadings and, also, offers a different point of view for interpreting the rotated components. A great many different rotation criteria exist which can be applied to the coefficients scaled in various ways. So, often the best interpretation is obtained by trial and error. However, as always the case when thresholding is applied, the non zero loadings are not the same as they would be if the loadings disregarded were zero. Therefore the analysis is somewhat inaccurate.
%
%

%\centerline{======== SPCA ===========}
\subsection{Sparse principal component analysis}
Because of the drawbacks of thresholding the coefficients, direct methods that aim to find different sets of components with sparse coefficients, called sparse PCA (SPCA), have been a subject of extensive study over the past few years.
The first of these method aimed at simultaneously maximise the components' variance (Hotelling's formulation of PCA) and the Varimax rotation method \cite{jol00}. Even though, as its authors also point out, this method has some drawbacks, it was a breakthrough. In fact, it started a stream of SPCA methods \cite[among others][]{jol00, jol03, zou, mog, sri, jou} which maximise the variance of the components with the addition of a different sparsity penalty. These methods compute the sparse components as the first pcs of different subsets of variables, or \emph{blocks} of variables, \citep{mog} and also have some drawbacks, including not maximising the variance explained by the components \cite[see][for a dicussion]{mer15, mer19}.%

Least square SPCA methods \cite[LS SPCA,][]{mer15}, differently from conventional SPCA methods, compute sparse components by maximising the variance that they explain. Projection SPCA \cite[PSPCA,][]{mer19} is a variant of LS SPCA with which sparse components are computed by regressing (projecting) the pcs onto subsets of variables. It can be shown that the proportion of variance explained retained by these components is not less than the coefficient of determination of the regression.

%%%%%%%%%%%%%%%%%%%%  SIMPCA ============================================
%============== SIMPCA ===============================
\subsection{SIMPCA}
In this paper we propose SIMPCA, a method for computing rotated sparse components.
The basic idea is to apply PSPCA to rotated pcs. This means that, after rotation, the pcs are simplified by regressing each of them onto a subset of variables. The variables onto which the rotated pcs are projected can be chosen either by thresholding their coefficients or by using a regression variable selection strategy.

The motivation for SIMPCA is to replace the practice of thresholding rotated pcs coefficients with a method that produces components with some truly zero coefficients and explain a large proportion of the original full cardinality rotated pc. The choice of using PSPCA, instead of other SPCA methods, is due to the need of maintaining the direction determined by the rotation unchanged. The algorithm is quite simple to implement, as it only requires libraries for singular value decomposition or eigen-analysis and modest matrix computation skills. It can also be run ``manually'' using a software which rotates the pcs and performs regression variable selection.

The SIMPCA framework can accommodate different rotation and variable selection criteria. Through changing these, different sparse solutions can be found an evaluated. Merits and demerits of rotating the PCs are not addressed by this method, which simply provides truly sparse rotated principal components. We will show that useful simplifications can be easily obtained with SIMPCA. Furthermore, when discussing the rotation of pcs coefficients we show that any orthogonal rotation applied to the coefficients scaled to unit $\Lt$ norm are equivalent.

The paper is organised as follows: in the next section we present the SIMPCA methodology. In Section \ref{sec:simpca} we describe the SIMPCA algorithm and discuss rotation and variable selection criteria. In Section \ref{sec:applications} we illustrate the method with numerical examples based on real data sets. Finally, in Section \ref{sec:remarks} we give some concluding remarks.

%%%%  SECTION SIMPCA ===================================================================
\section{Overview of methods used for sparsifying principal components}\label{sec:simpca}
In the following we assume that the $n$ observations on $p$ variables have been centred to zero mean and stacked one over the other in the $(n\times p)$ matrix $\bX$. We also assume that the variables have been scaled so that  $\bS = \bXT\bX$ is equal to the covariance or correlation matrix.

We presume that the readers are familiar with standard PCA theory \citep[e.g.][Ch 7.2]{ize}. So, it should be understood that PCA is intimately related to the singular value decomposition of the matrix $\bX$, for which $\bX = \bU\bLa\bV\trasp$. So that, the coefficients (vectors) satisfy $\bX\trasp\bX\bv_j = \bv_j\lambda_j$  and the pcs are equal to
\begin{equation}
\bP_{[d]} = \bU_\subd\bLa_\subd = \bX\bV_\subd,
\end{equation}
where the subscript $\subd$ denotes the first $d$ columns of a matrix. For simplicity of exposition, this subscript will only be shown when it is necessary to specify the number of components considered.

Finding sparse pcs entails two distinct problems:
\begin{enumerate}
\item identify the indices of the variables corresponding to non-zero coefficients;
\item compute the non-zero coefficients once the indices have been identified.
\end{enumerate}
In the following we give a brief overview of existing methods used to sparsify principal components. Further discussion will be given in later sections.
\subsection{Thresholding the coefficients}\label{sec:thresh}
Thresholding is the oldest and simplest method used for simplifying the pcs. With this method,
the nonzero coefficients of each sparse component are set to be equal to the coefficients of the corresponding pc with absolute value larger than a prechosen threshold value. Hence, thresholding only identifies the indices of the nonzero coefficients without recomputing them, even though, sometimes, these coefficients are rescaled to have $\Lt$ norm equal to one.

Thresholding the coefficients can give misleading results for the following reasons:
\begin{enumerate}
\item The choice of the threshold value is subjective and the effects of the choice are not known. In most cases the same threshold value is used for all simplified pcs;

Traditionally unit $\Lt$ norm coefficients are thresholded but the coefficients can be scaled to different norms prior of thresholding. Jeffers \cite{jef} recommended scaling the coefficients to unit $\Lmax$ norm, that is dividing them by their maximum absolute value. Another useful choice is the unit $\Li$ norm scaling. This scaling transforms the coefficients to percent \emph{contributions} which makes their values and the threshold chosen more meaningful.

Normalising the coefficients with respect to a norm $\text{L}_m =\left(\sum_{i = 1}^{p}|a_i|^m\right)^{1/m}$ increases the divergence between ``large'' and ``small'' values as the value $m$ increases. The choice of the threshold value is crucial for the simplification obtained. However, it should be considered that normalisation is a nonlinear but monotonic transformation, and the same thresholding result is obtained by applying different threshold values to the coefficients scaled with respect to different norms. Hence the scaling is useful only to help choosing a meaningful threshold. The plot in Figure \ref{fig:exnorms} shows the effect of normalising the coefficients with respect to different norms.
 \begin{figure}[H]
    \centering
        \includegraphics[width = 0.5\textwidth]{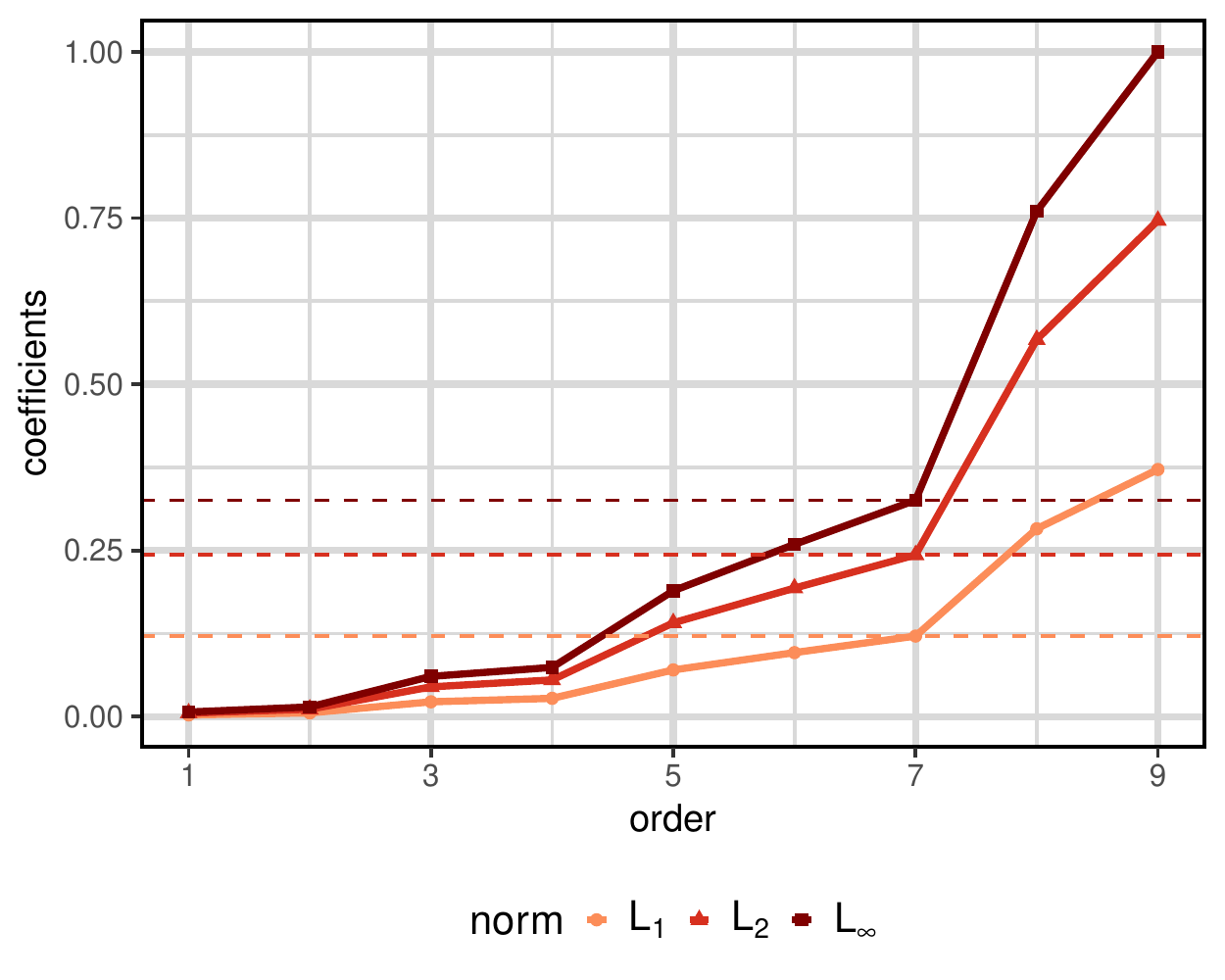}
    \caption{Coefficients scaled to different norms. The broken lines show threshold values which determine the same result for different scalings.}
    \label{fig:exnorms}
 \end{figure}
In the literature different rules of thumb are suggested for choosing a threshold value. In general, small variations of the threshold will not change the result. Choosing too large a threshold may lead to a failure of the algorithm because all coefficients are set to zero. The more variables in the set and the smaller the threshold should be. In fact, for $p$ variables, when the coefficients are scaled to norm $\text{L}_m$, there must be at least one coefficient per column not smaller than $p^{-1/m}$. So, the lower bounds for the largest coefficients are $1/p$ or $1/\sqrt p$ or one when the coefficients are scaled to $\Li$ or $\Lt$ or $\Lmax$, respectively. These values can be used as reference for how strong a simplification is defined by a threshold. In some cases we had good results with thresholds higher than these limits.

We found that iteratively including the variables corresponding to the larger coefficients until a certain amount of variance explained is reached, separately for each component, leads to more satisfactory solutions than using a fixed threshold value for all components. The extra variance explained by the sparse components can be computed with the formulae given in \cite{mer19}. We will give examples of this iterative reverse thresholding method procedure in later sections.
%                      \emph{contributions} and it is often used, for example, in Statistical Process Control (\eg, see \citealp{mer}). For unit length coefficients most orthogonal rotations are equivalent, as we will show in Section \ref{sec:rot}.

\item The variables selected with thresholding are generally highly mutually correlated. This can be seen by considering that a pc's coefficients are proportional to the correlations of the variables with the pc. In fact, for a component $\bp_j = \bX\bv_j$,
    \begin{equation*}\label{eq:corxpc}
      \bv_j = \frac{\bX\trasp\bX\bv_j}{\lambda_j} = \frac{\bX\trasp\bp_j}{\lambda_j} = \left\{\frac{\bx\trasps{i}\bp_j}{\lambda_j}\right\}_{i=1}^p =
      \left\{\text{corr}(\bx_i, \bp_j)\sqrt{\frac{\bx\trasps{i}\bx_i}{\lambda_j}}\right\}_{i=1}^p.
    \end{equation*}
%($v_{ij} \propto \text{corr}(\bx_i, \bp_j)s_{ii}$, where $s_{ii}$ is the standard deviation of$\bx_i$).
When PCA is carried out on the variables standardised to equal norm ($\bx\trasps{i}\bx_i = \bx\trasps{j}\bx_j,\forall i, j$), the coefficients are proportional to the correlations. It is well known that variables highly correlated with a common variable are also highly correlated among themselves (for example, see \citep{mer19} Lemma 1, for a proof). In some cases the variables selected with thresholding could be multicollinear.
%
% in most cases variables corresponding to the nonzero coefficients correspond to highly correlated variables. This can be seen by considering that the pcs' coefficients are proportional to the correlation of the corresponding variable with the pc multiplied by the variable standard deviation. In fact,
% When the pcs are computed on the correlation matrix the coefficients are equal to the correlations
% $\text{corr}(\bx_i, \bp_j)= \frac{\bx\trasps{i}\bp_j}{\sqrt{\bx\trasps{i}\bx_i\lambda_j}}$. In general, .
%
\item
Thresholded components are no longer mutually orthogonal and the variance that they explain is no longer equal to their norm. In some cases, thresholded components could explain less variance than thresholded components of higher order. %to see this, let $\dvj$ be the vector of coefficients
\end{enumerate}
\subsection{Conventional SPCA methods}
Thresholding does not yield truly parsimonious solutions because ``small'' coefficients are simply ignored. Consequently,  the properties of the so sparsified components are not foreseeable. For this reason, researchers sought alternative methods which would yield genuinely parsimonious solutions by optimising a criterion under the constraint that some of the coefficient are equal to zero. A great many SPCA methods derived from adding a sparsity requirement to Hotelling's formulation of the PCA problem (right-hand side term of Equation \ref{eq:pcaopt}) have been proposed.

With this approach, the sparse components are computed as the first pcs of different subsets of variables\footnote{This is true also for methods derived from a psuedo least squares optimization (such as SPCA \citep{zou}, for example) because of a constraint added to the least squares function \citep[see][for a proof]{mer15}.} \citep{mog}. If we let $\daj$ be the set of nonzero coefficients of the $j$-th sparse components, corresponding to predetermined indices, and $\dXj$ the corresponding subset of variables, the optimal coefficients are determined as
\[
\daj = \argmax\limits_{\dc\trasps{j}\dcj = 1}\dc\trasps{j}\dX\trasps{j}\dXj\dcj + \phi(\dcj),
\]
where $\phi(\dcj)$ is a penalty function. The indices of the nonzero coefficients for each component are determined under different constraints with sophisticated greedy algorithms, designed to be efficient on large matrices. Merola \cite{mer15, mer19} shows that the solutions provided by these methods are similar to those obtained with thesholding and suffer from similar drawbacks, including that:
\begin{enumerate}
  \item the sparse components are derived by assuming that their norm is equal to the variance that they explain, which is not true;
  \item the nonzero coefficients correspond to highly correlated components;
  \item in most cases the components are not orthogonal;
  \item in some cases the components are defined as linear combinations of differently defined residuals of the original variables. Hence, the components computed on the original variables have different properties;
  \item computing the solutions requires complex algorithms and setting obscure parameters;
  \item when some of the original variables are perfectly correlated, these methods fail to find the sparse representation of the standard pcs \cite{mer19}.
\end{enumerate}
\subsection{Least squares  sparse PCA (LS SPCA)}
In LS SPCA \cite{mer15} the sparse components are computed by adding a sparsity constraint to Pearson's minimisation of the approximation error (left-hand side term of Equation \ref{eq:pcaopt}). Hence, for a given set of indices, the nonzero coefficients of the $j$-th LS SPCA component, $\daj$, are the solution to
\begin{equation}\label{eq:lsspca}
  \daj = \argmin \vert\vert \bX - \dXj\daj\bl\trasps{j}\vert\vert^2 =
  \argmax \frac{\dajT\dXjT\bX\bX\trasp\dXj\daj}{\dajT\dXjT\dXj\daj}.
\end{equation}
The LS SPCA solutions sequentially maximise the extra variance explained by the each sparse component, $\bt_j = \dXj\daj$, which is defined as
\begin{equation}\label{eq:lsspcavexp}
  \vexp(\bt_j) = \frac{\bt\trasps{j}\bQ\trasps{j}\bQ_j\bt_j}{\dajT\dQjT\dQj\daj},
\end{equation}
where $\bQ_j$ denotes the orthocomplement of the $\bX$ matrix with respect to the previously computed components ($\bQ_1 = \bX$). The solutions are then obtained as generalised eigenvectors either requiring that the components are mutually orthogonal (USPCA) or simply that the extra variance explained is maximised (hence dropping the orthogonality requirement, CSPCA).

Finding the globally optimal LS SPCA  indices of the nonzero coefficients is an NP-hard non-convex problem, hence computationally intractable (as is for conventional SPCA,\cite{mog}). Merola \cite{mer15}) proposes to determine the indices for each component by requiring that the resulting sparse components explain at least a proportion $\alpha$ of the variance explained by the corresponding pc. The greedy algorithms proposed to find the indices with this criterion (a branch and bound and backward iterative elimination) give excellent results but are not efficient on large matrices.

\subsection{Projection Sparse PCA (PSPCA)}
Merola \cite{mer19} proposes to compute sparse pcs by simply projecting each pc, $\bp_j = \bX\bv_j$, onto a block of variables, $\dXj$. Hence, the PSPCA sparse components are defined as
\begin{equation}\label{eq:pspcacomp} %
  \hpj = \dXj(\dXjT\dXj)\matinv\dXjT\bpj = \dXj\dvj,
\end{equation}
where $\dvj = (\dXjT\dXj)\matinv\dXjT\bpj$ denotes the nonzero coefficients.
It can be proved that the proportion of total variance of $\bX$ explained by this projection is
not smaller than the coefficient of determination of the regression. That is, $\vexp(\hpj) \geq \alpha\,\vexp(\bpj)$, where $0 < \alpha \leq 1$ is the coefficient of determination of the regression. Since the pcs are linear combinations of the variables in $\bX$, it is always possible to select a subset of variables, $\dXj$, to ensure that the projection will yield a coefficient of determination not less than any $\alpha \leq 1$.
This means also that the correlation between the PSPCA components and the full cardinality pcs is such that $\vert corr(\bpj, \hpj)\vert =\sqrt{\alpha}\geq \alpha$.

If more than one sparse component are required, the process can be repeated for a set of full cardinality components. In this case, recomputing the pcs from the matrix of orthogonal residuals, $\bQ_j$, after each sparse component is computed, yields better results. When the original components are uncorrelated, pairs of projected components will not be uncorrelated but their squared correlation will be not greater than $1 - \alpha$, hence very small for large $\alpha$ \cite{mer19}.
\subsection{Variable selection}
PSPCA can be used as a variable selection method for computing LS SPCA components.
In fact, it can be shown that LS SPCA components explain at least as much variance as PSPCA components computed on the same subset of variables. This means that the variables for LS SPCA sparse components can be efficiently selected with one of the variable selection algorithms used in regression using the PCs as response variables.

In this section we illustrate some of the differences between sparse pcs computed by thresholding the coefficients and LS SPCA sparse pcs obtained on variables selected with forward regression selection. For thresholding we consider both the reverse iterative procedure (Section \ref{sec:thresh}, Point (1)) and using the same threshold value for all components. These are compared to the corresponding sparse components computed with both USPCA and CSPCA.

The \emph{Costa Rican Household Poverty Level Prediction} dataset was used for a Kaggle competition\footnote{Data and their description can be found at \url{https://www.kaggle.com/c/costa-rican-household-poverty-prediction/overview}}. We selected only the records relative to household heads in the training sample, removing those with missing values. The final dataset contained a total of 120 socio-economic measures on 2970 households, which were centered  to zero mean and scaled to unit variance because heterogenous.

\begin{figure}[H]
   \centering
        \includegraphics[width = 0.33\textwidth]{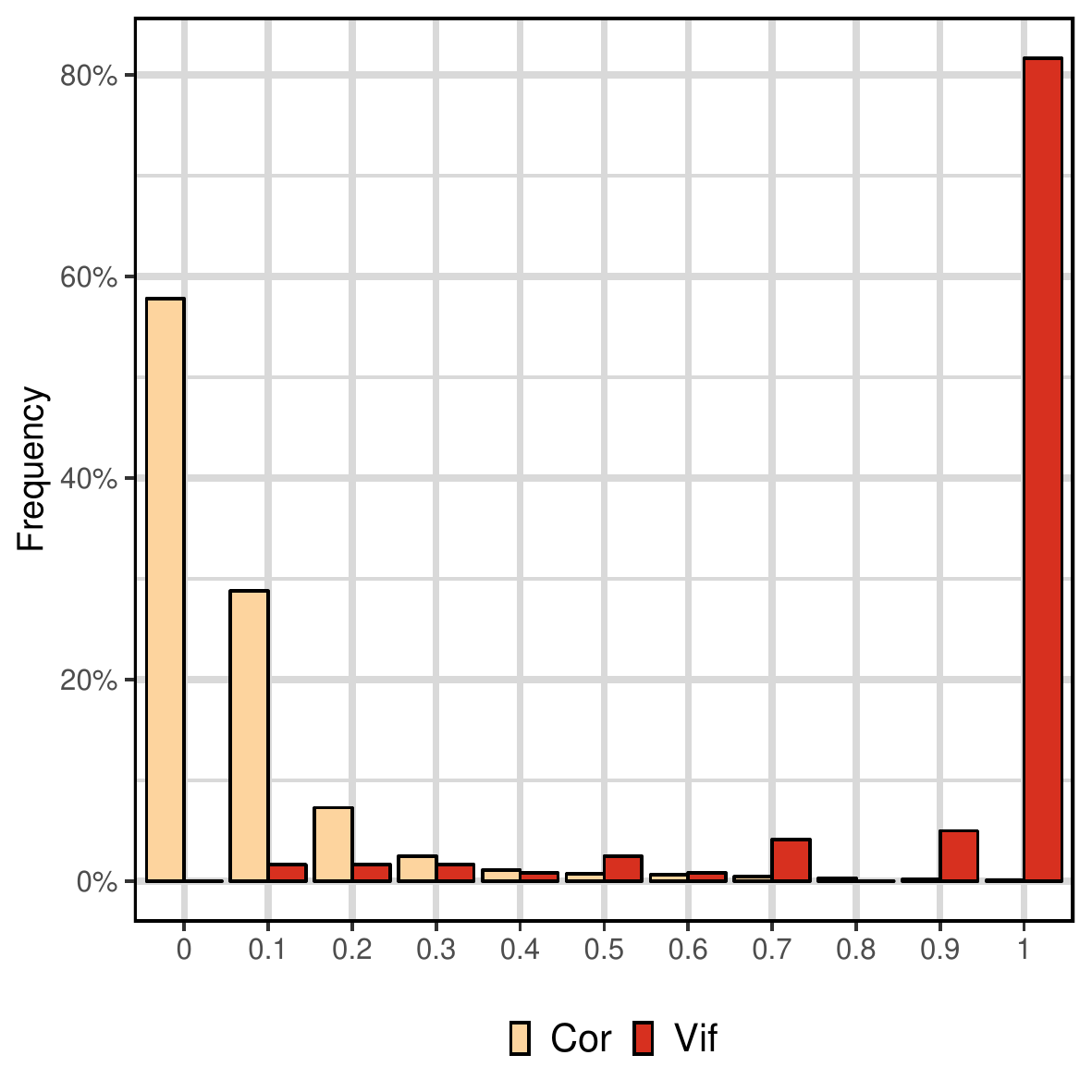}
    \caption{Distribution of the pairwise (in absolute value) and squared multiple correlations between variables in the Costa Rica dataset.}
    \label{fig:ch_corvif}
\end{figure}

While most of the variables in the dataset have low pairwise correlation, many of these have high or perfect multiple correlation with the others. This can be appreciated by observing Figure \ref{fig:ch_corvif}, which shows the distribution of the pairwise correlations (in absolute value, Cor) and the squared multiple correlations (measured by the coefficients of determination of regressing each variable on all others, also known as variance inflation factors, Vif).

\begin{figure}[H]
   \centering
        \includegraphics[width = 0.5\textwidth]{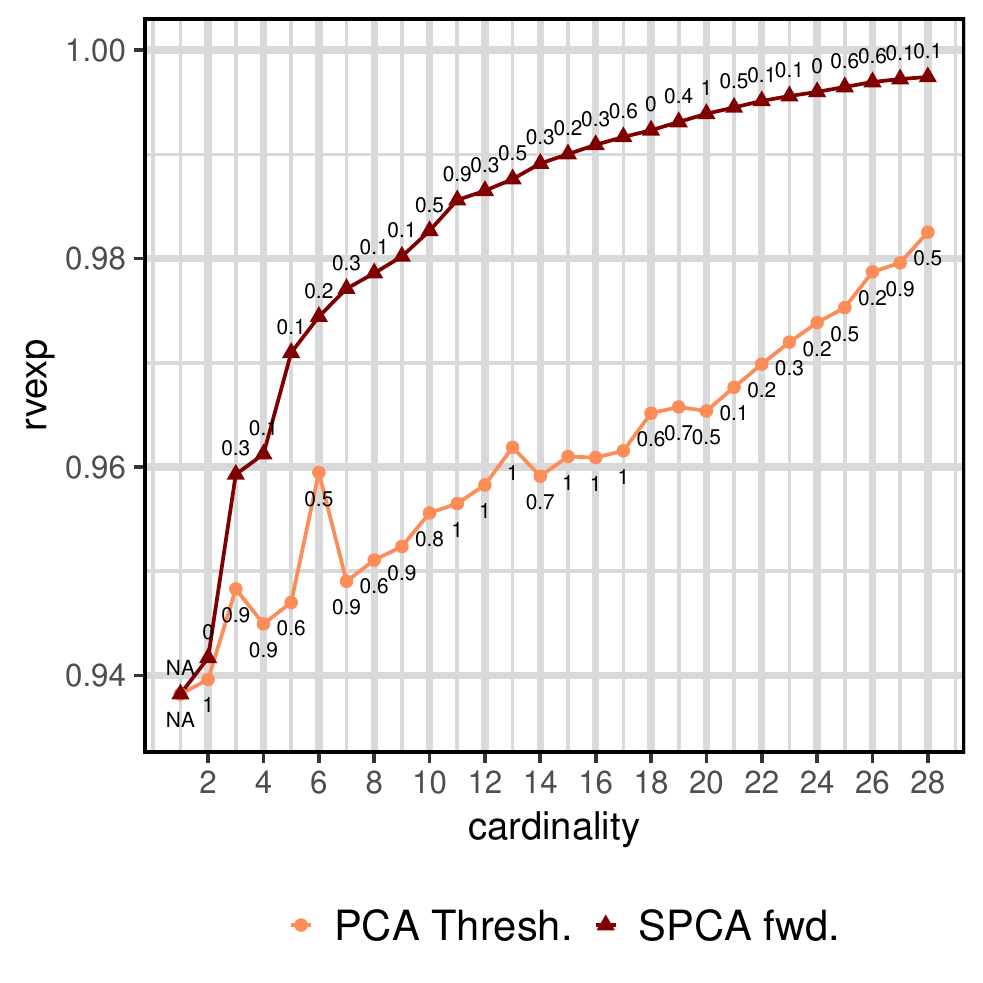}
    \caption{Relative variance explained by iteratively selecting the variables for the first components. The labels denote the squared multiple correlation of the newly entered variable with the ones already selected.}
    \label{fig:ch_iterselect}
\end{figure}

\begin{figure}[H]
   \centering
        \includegraphics[width = 0.66\textwidth]{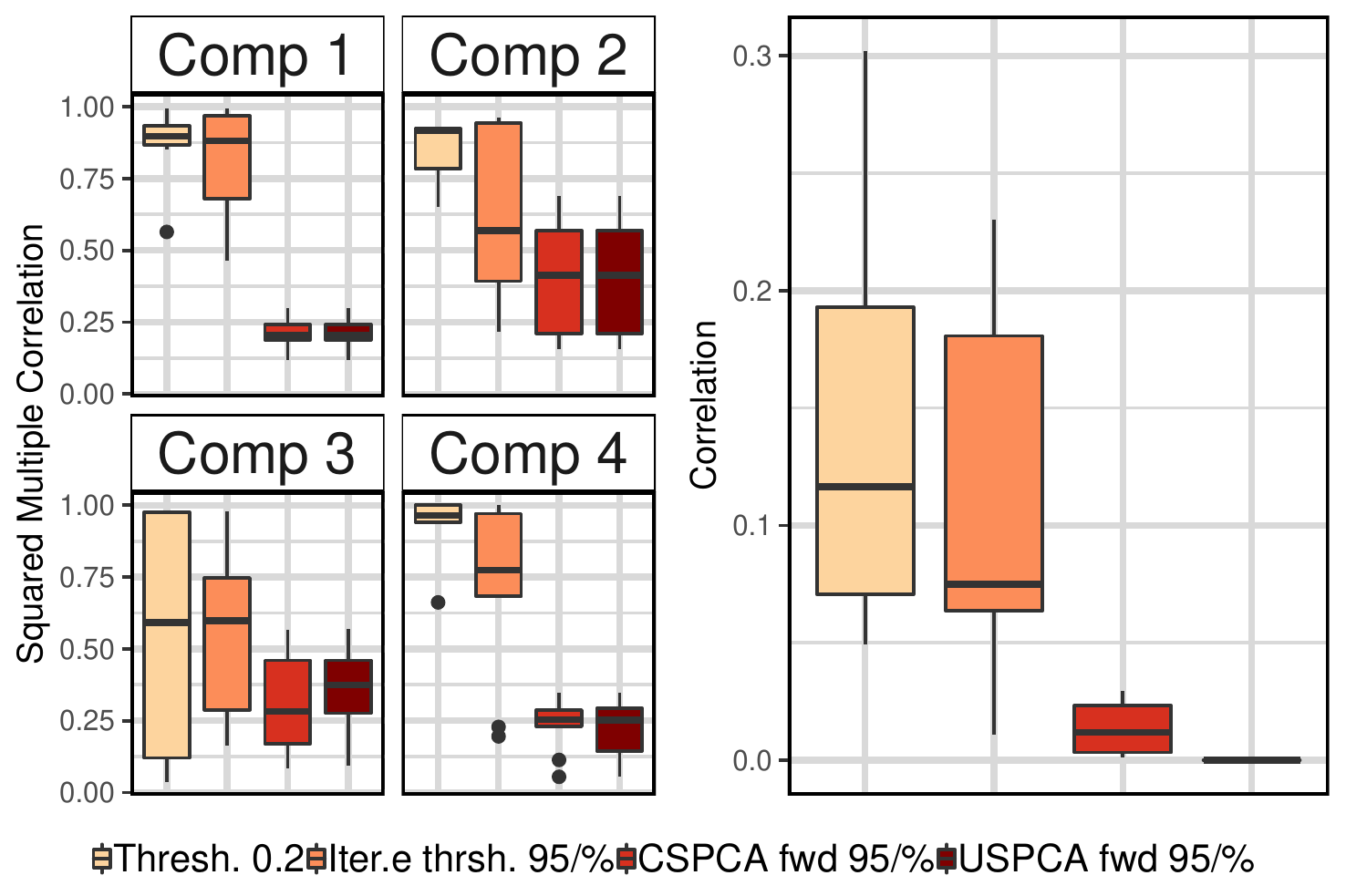}
    \caption{Distribution of the variance inflation factors within the variables selected for the first four sparse components (left) and correlation between components (right) for different selection methods.}
    \label{fig:ch_vifCor4comps}
\end{figure}

Figure \ref{fig:ch_iterselect} shows the variance explained, relative to that explained by the first pc (rvexp), by the first thresholded and USPCA components.
The variables were iteratively selected by reverse thresholding and forward selection (up to the number of variables, 28, taken by thresholding to reach 98\% relative variance explained). The labels show the squared multiple correlation of each newly selected variable with the variables previously selected.
The USPCA component always explains more variance than the thresholded one and reaches 98\% rvexp with 10 variables whereas thresholding takes 28. As expected, many of the variables selected with thresholding  are highly multiply correlated, especially the ones corresponding to larger coefficients. It should be observed that, since thresholded components are not recomputed, adding new variables in some cases determines a drop in variance explained.
We computed four different sets of four sparse components: thresholding the pcs with fixed threshold value equal to 0.2 and iterative reverse thresholding with $\alpha = 0.95$, and computing the USPCA and CSPCA components with forward selection, also with $\alpha = 0.95$.
Figure \ref{fig:ch_vifCor4comps} shows the distributions of the Vifs among the variables selected and the correlation between sparse components obtained for the different sets of components.
It can be appreciated how thresholded components always include more correlated variables and the corresponding components are more correlated than the LS SPCA ones.

% applying PSPCA to the rotated pcs.
% In the original PSPCA algorithm the components to be approximated is computed after the previous component sparse approximation is obtained in order to recover part of the variance explained lost in the approximation. Since in this context the aim is to preserve the rotated structure, recomputing the components is not  necessary.

\subsection{Rotation of the pcs coefficients}\label{sec:rot}
The several existing rotation criteria are determined by maximising some measure of simplicity (or minimising a measure of complexity) of the coefficients\footnote{Rotations were originally introduced to rotate the loadings in Factor Analysis, a DRM discussing which is outside the scope of this paper. In the following we refer only to rotations of the pcs coefficients.}. Simplicity has been defined in different ways. Thurstone's \cite{thu} widely cited rules for simple structure for a matrix with $m$ columns are:
\begin{enumerate}
\item Each row should contain at least one zero. [\emph{parsimony}, row-simplicity]
\item Each column should contain at least $m$ zeros. [\emph{sparseness}, column-simplicity]
\item Every pair of columns should have several rows with a zero in one column but not the other.
\item If $m\geq 4$, every pair of columns should have several rows with zeros in both columns.
\item Every pair of columns should have few rows with nonzero coefficients in both columns.
\end{enumerate}
Obviously, these rules cannot be all simultaneously included in an objective function and several different simplicity criteria have been proposed \cite[see][for an excellent overview]{bro}. A comprehensive family of complexity functions for the rotations of coefficients is the Crawford-Ferguson criterion \cite[CF,][]{cra}), shown in Equation (\ref{eq:cf}). This generic measure is a combination of a measure of sparseness and a measure of parsimony. By changing the parameter $\kappa$, almost all of the existing rotation criteria can be obtained. Orthogonal and oblique CF solutions can be computed using a derivative free algorithm \cite{jen, jen02}. Orthogonal CF criteria are equivalent to the Orthomax criteria \cite{har}, shown in Equation (\ref{eq:orthomax}).

Rotations methods were designed to rotate components with unit variance ($\Lt$ norm in the sample). Instead, when these are applied to standard pcs they give different results \cite{jol95}, because each pc has variance equal to the corresponding eigenvalue. It should be noticed that, in some cases, the coefficients are scaled to unit $\Lt$ row norm prior to rotation (Kaiser normalisation \cite{kai}). In some case this procedure gives better results but it changes the variance of the components.

Orthogonal rotations yield orthogonal rotated pcs only if applied to the coefficients scaled to produce components with equal $\Lt$ norm, $\bA = g\bV\bLa^{-1/2}$, for some $g\neq 0$, where $\bI$ is the Identity matrix. This follows from the definition of orthogonal rotation as a $d\times d$ orthonormal matrix, $\bO$, such that $\bO\trasp\bO = \bO\bO\trasp = \bI$, as per Equation (\ref{eq:pcamodrot}). A set of rotated pcs, $\btP = \bX\bV\bO$ will be orthogonal if $\bO\trasp\bP\trasp\bP\bO = g\bI$, which is verified only if the pcs, $\bP = \bX\bV$, are such that $\bP\trasp\bP = g\bI$.

An interesting property of rotations applied to unit $\Lt$  pcs coefficients is that all optimal orthogonal CF rotation criteria will yield the same solution, as we prove next.
% for a matrix with unit $\Lt$ columns, orthogonal rotations minimising any CF criterion are equivalent. Hence, anyone of these will yield the same results as Orthomax  criteria, including the frequently used Varimax and Quartimax. This concerns us because it applies to the rotation of unit length pcs coefficients.

\newtheorem{Thcf}{Theorem}
\begin{Thcf}
    Let $\bA$ be a $p\times d$ matrix with orthogonal columns, such that $\bA\trasp\bA = g\bI$, for some scalar $g>0$. Let also  $\bO$ be the orthonormal rotation for which $\bB = \bA\bO$ minimises the CF criterion
    \begin{equation}\label{eq:cf}
        cf(\bB; \kappa) = (1-\kappa) \sum_{i=1}^p\sum_{j=1}^d\sum_{k\neq j}^d b_{ij}^2b_{ik}^2 +
        \kappa \sum_{j=1}^d\sum_{i=1}^p\sum_{k\neq i }^p b_{ij}^2b_{kj}^2.
    \end{equation}
    For any choice of $0\leq \kappa\leq 1$, the minimisation of the orthogonal CF criterion is equivalent to the maximisation of the Quartimax criterion
    \begin{equation*}
       p \sum_{i=1}^p\sum_{j=1}^d b_{ij}^4,
    \end{equation*}
which has  a unique optimal solution.
   \begin{proof}
    The minimisation of the orthogonal CF criterion with any value $\kappa$  is equivalent to maximisation of the Orthomax criterion with $c = p\kappa$ \cite[][pp. 324-325]{cra}
    \begin{equation}\label{eq:orthomax}
        O(\bB; c) = p \sum_{i=1}^p\sum_{j=1}^d b_{ij}^4 +
        c\; \sum_{j=1}^d\left(\sum_{i=1}^p b_{ij}^2\right)^2.
    \end{equation}
Since $\bA\trasp\bA =  g\bI_d$ and $\bO\trasp\bO = \bI_d$, then $\bB\trasp\bB = g\bI_d$. This implies that $\sum_{i=1}^p b_{ij}^2 = g, \forall\, j$. Substituting this and $p\kappa$ for $c$ in Equation (\ref{eq:orthomax}), the Orthomax criterion reduces to:
        \begin{equation*}
            O(\bB; p\kappa) = p \sum_{i=1}^p\sum_{j=1}^d b_{ij}^4 +
            g\,d\,p\kappa.
        \end{equation*}
Since $\bB$ does not depend on $g, d, p$ and $\kappa$, $O(\bB; p\kappa)$ is equivalent to the Quartimax criterion. The theorem is proved.
    \end{proof}
\end{Thcf}
This theorem is important because the orthogonal CF criterion includes several known criteria. We have the following:
\newtheorem{Cocf}{Corollary}
\begin{Cocf}Any orthogonal rotation of equal length coefficients that minimises a Crawford-Ferguson criterion also maximise any Orthomax criterion, including the Quartimax, Varimax and Equamax criteria, among others.
\end{Cocf}
\section{SIMPCA}\label{sec:algo}
SIMPCA wants to be a methodology for sparsifying rotated pcs in which standard thresholding is replaced by PSPCA. Since there are no prescribed criteria for enhancing interpretability, SIMPCA can be applied under different settings, such as different rotation and variable selection criteria.

Since we aim to approximate rotated pcs, which no longer explain the most possible variance, PSPCA is the preferred method for computing the sparse components. This is because the projected components are the best sparse approximation (in a least squares sense) of the rotated pcs that can be computed from a block of variables. Instead, other methods may find sparse components which explain more variance but will necessarily depart from the direction of the rotated components.

The SIMPCA algorithm, outlined in Algorithm \ref{algo:simpca}, is rather simple because it only wants to automatise a number of common operations. In the initialisation, Step (0), we
included some possible options but more can be used. A more sophisticated version of the algorithm may include recomputing the pcs from the orthocomplement of the previously computed components, as done in PSPCA, which are then rotated. For simplicity of exposition, we do not show this variation, which is simple to implement if desired.
%These include rotation criterion, scaling of the coefficients, number of  and selection of the variables.
\begin{algorithm}[H]%[h]
\caption{SIMPCA algorithm}
\footnotesize{
\noindent
\begin{enumerate}[label = \arabic*:, start = 0, align = right, leftmargin=*]
   \item  \emph{Initialisation: before running SIMPCA there are different choices to be made:}
  \begin{enumerate}[label=0.\arabic*:]%[label= {0}.\arabic*.]
     \item \emph{the number of dimensions to retain}, $nd$.
     \item \emph{the scaling of the coefficients}: $SCALE$;
     \item \emph{the number of components to rotate at each step, $nr$,and the rotation method}: $ROTATE$;
      \item \emph{the variable selection criterion}: $SELECT$;
  \end{enumerate}
\end{enumerate}
\begin{algorithmic}[1]
 \Procedure {SIMPCA}{}
    \State{$\bV \gets PCA(\bX)$} \Comment{Compute coefficients principal components}
   \If{(scale = true)}\Comment{if required scale coefficients}
        \State{$\bV_{[nr]}  \gets SCALE(\bV_{[nr]})$}
    \EndIf
    \State{$\btV \gets ROTATE(\bV_{[nr]})$}\Comment{rotate $nr$ coefficients}
    \State{$\btP \gets \bX\btV$}\Comment{compute rotated pcs}
    \For{$(j = 1\, \mathbf{to}\, nd)$} \Comment{\textbf{sparse components computation}}\label{algo:loop}
    \State{$\dXj \gets SELECT(\bv_j)$}\Comment{select the block of variables}
    \State{$\dwj \gets (\dXjT\dXj)\matinv\dXjT\btp_{j}$}\Comment{compute the sparse coefficients}
    \State{$\brj \gets \dXj\dwj$}\Comment{compute the $j$-th sparse rotated component}
     \EndFor\Comment{\textbf{end components computation}}%
 \EndProcedure
  \end{algorithmic}
  }
  \label{algo:simpca}
\end{algorithm}

\section{Application examples}\label{sec:applications}
%\subsection{Numerical comparisons}\label{sec:numcomp}
%\subsection{Examples}
In this section we illustrate the application of SIMPCA to various datasets freely available on the Web. Details and references to these datasets can be found in Appendix \ref{app:data}. We chose these datasets because the measurements are simple to understand without specialised knowledge, and not because they gave the best results. We will briefly illustrate the application of SIMPCA to the data suggesting some interpretations but, by no means, carry out deep analyses of these datasets, which is beyond the scope of this paper and our expertise. All runs were made rotating the pcs coefficients so as to maximise the (orthogonal) Varimax criterion.

\subsection{Human Freedom}
To illustrate the differences between different variable selection methods, we use a collection of 79 distinct indicators of personal and economic freedom which was used to build the 2016 Human Freedom Index. For each of the 162 countries considered, the indicators describe the level of freedom on a scale from zero to 10 in the following areas:\\
\noindent
Peronal Freedom type: Rule of Law; Security and Safety; Movement; Religion; Association, Assembly, and Civil Society; Expression and Information; Identity and Relationships.\\
\noindent
Economic freedom type: Size of Government; Legal System and Property Rights; Access to Sound Money; Freedom to Trade Internationally; Regulation of Credit, Labor, and Business.

We removed the seven indicators relative to \emph{Association, Assembly, and Civil Society} because had a large number of missing values. Other missing values were generated by first imputing missing values in intermediate indices included in the dataset by Multivariate Imputations by Chained Equations \cite{vbu}. The imputed values of each index were then assigned to missing values of related variables. The final set had 74 indicators for 162 countries.

\begin{figure}[H]
\centering
    \includegraphics[width = 0.66\textwidth]{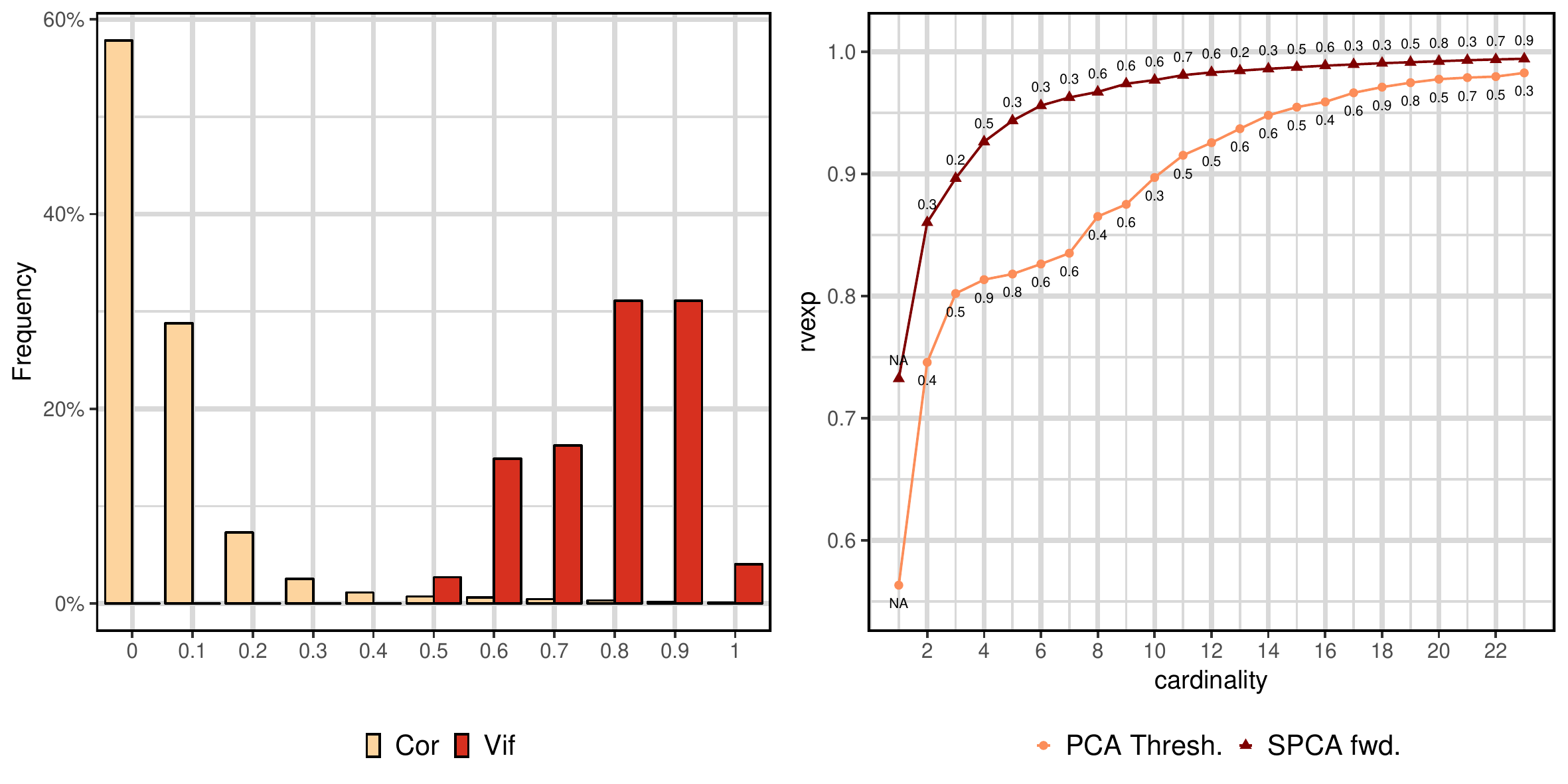}
\caption{Human freedom data. Distribution of the pairwise correlations (in absolute value) and squared multiple correlations between variables (Vifs) in the Human Freedom dataset (left). Relative variance explained by the variables in the first component selected iteratively with thresholding and forward selection (right). The labels denote the squared multiple correlations of the newly selected variables with those already in the model.}
\label{fig:hf_corvif_iter}
\end{figure}

Data were centred to zero mean but not scaled because the variables are homogeneous scores. The indicators have low pairwise correlation but about 65\% of them present multiple squared correlation not lower than 0.8, as shown in the left panel of Figure \ref{fig:hf_corvif_iter}.

We use this dataset to compare SIMPCA components computed using different variable selection strategies with those computed with standard thresholding, all derived from the coefficients of the first four pcs rotated with Varimax. We applied SIMPCA with $\alpha = 0.9$, recomputing the rotated pcs after each sparse component was determined. Variables were selected with forward, backward and (forward) stepwise selection algorithms, and adaptive thresholding. In adaptive thresholding we started with a threshold value of 0.25 for all components and decreased it by 0.05 if no coefficients in a set were higher than that value for a set of coefficients, until at least one coefficient was selected (the value was reduced to 0.2 only for the second component). For standard thresholding, that is setting to zero smaller coefficients without recomputing the others, we considered two strategies: iterative reverse thresholding, that is, adding one coefficient at the time starting with the largest, until the component explains 95\% of the variance of the corresponding rotated pc, and adaptive thresholding.

The plot on the right in Figure \ref{fig:hf_corvif_iter} shows the relative variance explained by
the first sparsified pc as 23 variables (the cardinality required to reach 95\% variance explained with iterative reverse thresholding) are added sequentially. The labels show the Vifs of the newly entered variables.
The first variables selected with forward selection have low multiple correlation and their inclusion quickly increases the proportion of variance explained, while the variables selected later determine a modest increase in the variance explained, because most of the variance of the pc has already been explained, and they are highly correlated with the previously selected variables. Conversely, the first variables selected with iterative reverse thresholding are highly correlated and determine a slower increase in variance explained.

Table \ref{tab:hf_summary} shows the summary statistics for the six sets of components obtained. The SIMPCA components obtained with a regression selection algorithm explain the variance more efficiently than those obtained with thresholding. This difference is more pronounced for the first two components (the ones that explain the most variance).
% Table generated by Excel2LaTeX from sheet 'AllSummaries'
\begin{table}[H]%[htbp]
  \centering
  \caption{Summary statistics for six types of sparsification of rotated components. rcvexp is the percentage of cumulative variance explained by the sparse components over that explained by the rotated pcs.}
  {\tiny
    \begin{tabular}{lrrrrrrrrrrrrrr}
     & \multicolumn{4}{c}{SIMPCA forward selection} &       & \multicolumn{4}{c}{SIMPCA backward  selection} &       & \multicolumn{4}{c}{SIMPCA stepwise  selection} \\
\cmidrule{1-5}\cmidrule{7-10}\cmidrule{12-15}          & \multicolumn{1}{l}{Comp1} & \multicolumn{1}{l}{Comp2} & \multicolumn{1}{l}{Comp3} & \multicolumn{1}{l}{Comp4} &       & \multicolumn{1}{l}{Comp1} & \multicolumn{1}{l}{Comp2} & \multicolumn{1}{l}{Comp3} & \multicolumn{1}{l}{Comp4} &       & \multicolumn{1}{l}{Comp1} & \multicolumn{1}{l}{Comp2} & \multicolumn{1}{l}{Comp3} & \multicolumn{1}{l}{Comp4} \\
vexp  & 24.5\% & 11.7\% & 5.5\% & 4.9\% &       & 23.4\% & 12.0\% & 6.0\% & 4.6\% &       & 23.4\% & 11.7\% & 6.0\% & 4.6\% \\
cvexp & 24.5\% & 36.3\% & 41.8\% & 46.7\% &       & 23.4\% & 35.4\% & 41.5\% & 46.1\% &       & 23.4\% & 35.2\% & 41.5\% & 46.1\% \\
rcvexp & 97.0\% & 94.0\% & 90.0\% & 90.0\% &       & 93.0\% & 92.0\% & 90.0\% & 89.0\% &       & 93.0\% & 92.0\% & 89.0\% & 89.0\% \\
card  & 3     & 2     & 2     & 3     &       & 2     & 4     & 2     & 2     &       & 2     & 2     & 2     & 2 \\
mincont & 25.5\% & 26.8\% & 32.0\% & 25.5\% &       & 42.7\% & 11.5\% & 32.0\% & 45.6\% &       & 42.7\% & 19.6\% & 32.0\% & 45.6\% \\
    \textcolor[rgb]{ 0,  0,  1}{} & \multicolumn{4}{c}{SIMPCA adaptive thresholding} &       & \multicolumn{4}{c}{Simple iterative thresholding} &       & \multicolumn{4}{c}{Simple adaptive thresholding} \\
\cmidrule{1-5}\cmidrule{7-10}\cmidrule{12-15}          & \multicolumn{1}{l}{Comp1} & \multicolumn{1}{l}{Comp2} & \multicolumn{1}{l}{Comp3} & \multicolumn{1}{l}{Comp4} &       & \multicolumn{1}{l}{Comp1} & \multicolumn{1}{l}{Comp2} & \multicolumn{1}{l}{Comp3} & \multicolumn{1}{l}{Comp4} &       & \multicolumn{1}{l}{Comp1} & \multicolumn{1}{l}{Comp2} & \multicolumn{1}{l}{Comp3} & \multicolumn{1}{l}{Comp4} \\
    vexp  & 24.9\% & 12.4\% & 7.4\% & 4.5\% &       & 22.8\% & 12.0\% & 7.3\% & 4.5\% &       & 24.8\% & 11.8\% & 7.2\% & 4.4\% \\
    cvexp & 24.9\% & 37.3\% & 44.7\% & 49.2\% &       & 22.8\% & 34.8\% & 42.1\% & 46.6\% &       & 24.8\% & 36.5\% & 43.8\% & 48.2\% \\
    rcvexp & 98.0\% & 97.0\% & 97.0\% & 95.0\% &       & 90.0\% & 91.0\% & 91.0\% & 90.0\% &       & 98.0\% & 95.0\% & 95.0\% & 93.0\% \\
    card  & 6     & 8     & 5     & 3     &       & 4     & 4     & 4     & 1     &       & 6     & 7     & 4     & 2 \\
    mincont & 12.8\% & 4.3\% & 12.6\% & 7.9\% &       & 20.1\% & 24.3\% & 20.1\% & 100.0\% &       & 13.1\% & 12.8\% & 20.1\% & 33.9\% \\
\cmidrule{1-5}\cmidrule{7-10}\cmidrule{12-15}    \end{tabular}%
}
  \label{tab:hf_summary}%
\end{table}%
Table \ref{tab:hf_contr1} shows the coefficients of the first sparse components computed with the different methods. All methods identify variables only of Personal Freedom type, specifically in Security and Safety Identity and Relationships areas. As expected, the variables selected with thresholding show larger multiple correlation than those selected with regression algorithms. The variables selected in SIMPCA with regression algorithms explain the variance more efficiently. For example, the cardinality three SIMPCA component obtained with forward selection explains almost the same proportion of variance as the cardinality six component obtained with adaptive thresholding. The component obtained with iterative reverse thresholding, with four variables explains less variance of the SIMPCA components of cardinality two.

Table \ref{tab:hf_contr2} shows the coefficients of the second sparse components computed with the different methods. Also in this case, regression selection algorithms determine components with lower cardinality which explain the variance more efficiently. All methods identify one variable of the Personal Freedom type (with the exception of iterative thresholding) and the others of the Economic Freedom type.
The two variables selected with forward selection are moderately correlated but they are of two different freedom types. Some of the variables selected by thresholding are highly correlated and concern the same area of freedom.
% Table generated by Excel2LaTeX from sheet 'contrib 1'
\begin{table}[H]%[htbp]
  \centering
  \caption{Human freedom data. Nonzero contributions of the first sparsified components computed with six different approaches.}
  {\tiny
     \begin{tabular}{lrrrrrr}
          & \multicolumn{1}{l}{SIMPCA fwd} & \multicolumn{1}{l}{SIMPCA bwd} & \multicolumn{1}{l}{SIMPCA step} & \multicolumn{1}{l}{SIMPCA thr} & \multicolumn{1}{l}{iter thr} & \multicolumn{1}{l}{adapt thr} \\
    \midrule
pf ss women inheritance widows & --    & --    & --    & 13\% (0.87) & 20\% (0.87) & 15\% (0.87) \\
pf ss women inheritance daughters & 38\% (0.56) & 57\% (0.4) & 57\% (0.4) & 19\% (0.89) & 20\% (0.88) & 15\% (0.89) \\
pf identity parental marriage & --    & --    & --    & 14\% (0.78) & --    & 14\% (0.78) \\
pf identity parental divorce & --    & --    & --    & 13\% (0.79) & --    & 13\% (0.79) \\
pf identity sex male & 37\% (0.41) & 43\% (0.4) & 43\% (0.4) & 26\% (0.64) & 35\% (0.64) & 25\% (0.64) \\
pf identity sex female & --    & --    & --    & 15\% (0.57) & 25\% (0.57) & 18\% (0.57) \\
pf identity divorce & 25\% (0.43) & --    & --    & --    & --    & -- \\
    \midrule
    card  & 3     & 2     & 2     & 6     & 4     & 6 \\
    rcvexp & 97.0\% & 93.0\% & 93.0\% & 98.0\% & 90.0\% & 98.0\% \\
    vexp  & 25.0\% & 23.0\% & 23.0\% & 25.0\% & 23.0\% & 25.0\% \\
    \bottomrule
    \end{tabular}%
    }
  \label{tab:hf_contr1}%
\end{table}%

% Table generated by Excel2LaTeX from sheet 'contrib 2'
\begin{table}[H]%[htbp]
  \centering
  \caption{Human freedom data. Nonzero contributions of the second sparsified components computed with six different approaches.}
 {\tiny
\begin{tabular}{lrrrrrr}
      & \multicolumn{1}{l}{SIMPCA fwd} & \multicolumn{1}{l}{SIMPCA bwd} & \multicolumn{1}{l}{SIMPCA step} & \multicolumn{1}{l}{SIMPCA thr} & \multicolumn{1}{l}{iter thr} & \multicolumn{1}{l}{adapt thr} \\
\cmidrule{1-7}
pf rol criminal & \multicolumn{1}{l}{73\% (0.67)} & \multicolumn{1}{l}{--} & \multicolumn{1}{l}{80\% (0.54)} & \multicolumn{1}{l}{--} & \multicolumn{1}{l}{--} & \multicolumn{1}{l}{--} \\
pf ss homicide & \multicolumn{1}{l}{--} & \multicolumn{1}{l}{11\% (0.11)} & \multicolumn{1}{l}{--} & \multicolumn{1}{l}{6\% (0.43)} & \multicolumn{1}{l}{--} & \multicolumn{1}{l}{14\% (0.42)} \\
ef legal judicial & \multicolumn{1}{l}{--} & \multicolumn{1}{l}{--} & \multicolumn{1}{l}{--} & \multicolumn{1}{l}{19\% (0.7)} & \multicolumn{1}{l}{25\% (0.65)} & \multicolumn{1}{l}{15\% (0.69)} \\
ef legal protection & \multicolumn{1}{l}{27\% (0.67)} & \multicolumn{1}{l}{51\% (0.47)} & \multicolumn{1}{l}{--} & \multicolumn{1}{l}{--} & \multicolumn{1}{l}{--} & \multicolumn{1}{l}{--} \\
ef legal military & \multicolumn{1}{l}{--} & \multicolumn{1}{l}{22\% (0.51)} & \multicolumn{1}{l}{--} & \multicolumn{1}{l}{16\% (0.6)} & \multicolumn{1}{l}{24\% (0.5)} & \multicolumn{1}{l}{14\% (0.59)} \\
ef legal integrity & \multicolumn{1}{l}{--} & \multicolumn{1}{l}{--} & \multicolumn{1}{l}{--} & \multicolumn{1}{l}{13\% (0.73)} & \multicolumn{1}{l}{25\% (0.67)} & \multicolumn{1}{l}{15\% (0.73)} \\
ef legal police & \multicolumn{1}{l}{--} & \multicolumn{1}{l}{--} & \multicolumn{1}{l}{--} & \multicolumn{1}{l}{17\% (0.75)} & \multicolumn{1}{l}{26\% (0.71)} & \multicolumn{1}{l}{15\% (0.73)} \\
ef trade regulatory compliance & \multicolumn{1}{l}{--} & \multicolumn{1}{l}{15\% (0.41)} & \multicolumn{1}{l}{--} & \multicolumn{1}{l}{10\% (0.59)} & \multicolumn{1}{l}{--} & \multicolumn{1}{l}{--} \\
ef regulation labor minwage & \multicolumn{1}{l}{--} & \multicolumn{1}{l}{--} & \multicolumn{1}{l}{--} & \multicolumn{1}{l}{4\% (0.2)} & \multicolumn{1}{l}{--} & \multicolumn{1}{l}{13\% (0.17)} \\
ef regulation business bureaucracy & \multicolumn{1}{l}{--} & \multicolumn{1}{l}{--} & \multicolumn{1}{l}{20\% (0.54)} & \multicolumn{1}{l}{15\% (0.64)} & \multicolumn{1}{l}{--} & \multicolumn{1}{l}{14\% (0.51)} \\
\midrule
card  & 2     & 4     & 2     & 8     & 4     & 7 \\
rcvexp & 94.0\% & 92.0\% & 92.0\% & 97.0\% & 91.0\% & 95.0\% \\
vexp  & 12.0\% & 12.0\% & 12.0\% & 12.0\% & 12.0\% & 12.0\% \\
\bottomrule
\end{tabular}%
}
  \label{tab:hf_contr2}%
\end{table}%

%% crime
\subsection{Crime data}
As a first illustration of SIMPCA, we compare the results of applying the method by thresholding the coefficients and forward selecting the variables to a dataset containing socioeconomic measures collected in the 1990s for different US cities. % (available at the UCI repository \cite{uci}).
All numeric data were normalized into the decimal range $0.00 - 1.00$ using an unsupervised, equal-interval binning method\footnote{In unsupervised equal-interval binning data are divided into k intervals of equal size. The width of the intervals is: $w = (\max - \min)/k$. The interval boundaries are:
   $\min, \min + w, \min + 2w, \cdots , \min + (k-1)w, \max$.}. Attributes retain their distribution and skew.

%\footnote{Communities and Crime Data Set,
%\url{https://archive.ics.uci.edu/ml/datasets/Communities+and+Crime}
The data were originally used to explain the rate of violent crime in the different cities. We deleted 22 variables with missing values and imputed by regression one missing observation. The final set contains 1994 observations on 100 variables. Our analysis was applied to the variables scaled to unit variance.

Before we discuss the results and their interpretation with respect to violent crime, we illustrate the difference between selecting the variables defining the sparse components by thresholding the coefficients and by forward selection. For a more comprehensive comparison of the methods, we show the results of selecting the variables for the first two standard pcs and the first two rotated pcs. Thresholding was applied with threholds equal to 0.15 for the pcs and 0.185 for the rotated pcs. Forward selection was applied requiring the coefficient of determination to be larger than 0.95 in all cases. The components were rotated by applying Varimax to the coefficients of the first four PCs.

Table \ref{tab:crimecompselectvar} shows cardinality, proportion of net variance explained by the projected components over the variance explained by the corresponding pc and median squared multiple correlation (Vif) among the variables selected for each component. Figure \ref{fig:crimebpvif} shows the distribution of the latter statistics for each case. It is evident that the variables selected by thresholding are much more highly correlated among themselves than the ones selected by forward selection. Hence, the sparse components produced with the latter selection method explain more variability with lower cardinality.
\spacingset{1}
\begin{table}[h]\label{tab:crimecompselectvar}
  \centering
  \caption{Crime data. Summary statistics comparing variable selection made by thresholding and  forward selection.}
{\scriptsize
    \begin{tabular}{lrrrrrrr}
    \toprule
          & \multicolumn{3}{c}{Threshold} &       & \multicolumn{3}{c}{Forward Selection} \\
\cmidrule{2-4}\cmidrule{6-8}    PCA   & \multicolumn{1}{l}{Cadinality} & \multicolumn{1}{l}{\% vexp} & \multicolumn{1}{l}{Median Vif} &       & \multicolumn{1}{l}{Cadinality} & \multicolumn{1}{l}{\% vexp} & \multicolumn{1}{l}{Median Vif} \\
    \midrule
    Comp 1 & 13    & 0.97  & 0.86  &       & 3     & 0.96  & 0.51 \\
    Comp 2 & 11    & 0.89  & 0.95  &       & 5     & 0.97  & 0.32 \\
    \midrule
    SIMPCA &       &       &       &       &       &       &  \\
    \midrule
    Comp 1 & 10    & 0.87  & 0.98  &       & 4     & 0.87  & 0.55 \\
    Comp 2 & 6     & 0.53  & 0.92  &       & 5     & 0.90   & 0.54 \\
    \bottomrule
    \end{tabular}%
}
  \label{tab:crimecompselectvar}%
\end{table}\spacingset{\bsln}%

\begin{figure}[h]\label{fig:crimebpvif}
  \centering
  \includegraphics[width=0.45\textwidth]{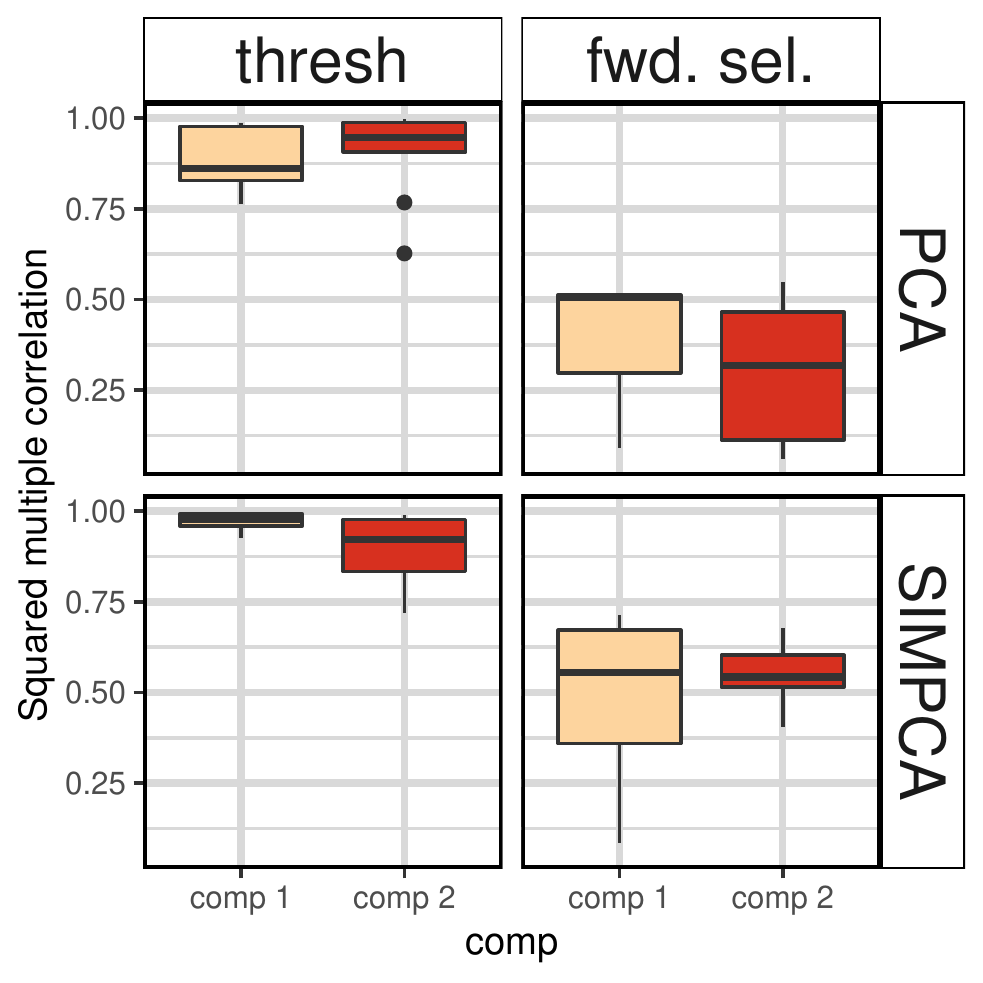}
  \caption{Crime data. Distribution of the multiple correlations among the variables selected with different methods.}
  \label{fig:crimebpvif}
\end{figure}%
The percent contributions to the first and second SIMPCA components are shown in Tables \ref{tab:crimecontrC1} and \ref{tab:crimecontrC2}, respectively. These tables also show the squared multiple correlation for each variable. The sign of the coefficients (the pcs are defined up to a change of sign) have been chosen so that the components have positive correlation with the rate of violent crime.

Of the ten variables selected by thresholding for the first component, seven concern housing and three income, confirming that they carry similar information. Instead, the variable selected by thresholding for the second component concern different aspects of the population and are more meaningful. However, it should be noted that ``number of people living in urban areas'' and ``population for community'' have 0.99 correlation but the contributions have different sign. Instead, the variables chosen with forward selection carry information on mainly different aspects of the population and have lower multiple correlations.
% Table generated by Excel2LaTeX from sheet 'Only Simpca'
\spacingset{1}
\begin{table}[h]
  \centering
  \caption{Crime data. Contributions to the first SIMPCA components. In parenthesis are shown the squared coefficients of multiple correlation.}
 {\scriptsize
    \begin{tabular}{lrr}
    \toprule
          & \multicolumn{2}{c}{comp 1} \\
\cmidrule{2-3}    Variable & \multicolumn{1}{l}{thresh} & \multicolumn{1}{l}{fwd. sel.} \\
    \midrule
    median family income & 16\% (0.93) & 45\% (0.71) \\
    median gross rent & -19\% (0.98) & -- \\
    number of people known to be foreign born & --    & 12\% (0.08) \\
    upper quartile owner occupied housing  & 3\% (0.99) & -- \\
    lower quartile owner occupied housing & 3\% (0.99) & -- \\
    median owner occupied housing & -3\% (1.00) & -- \\
    \% of people over 16 employed in manufacturing & --    & -13\% (0.45) \\
    per capita income & 13\% (0.97) & -- \\
    upper quartile rental housing & 21\% (0.97) & -- \\
    lower quartile rental housing & 19\% (0.94) & 30\% (0.66) \\
    median rental housing & 2\% (0.99) & -- \\
    per capita income for caucasians & -3\% (0.96) & -- \\
    \bottomrule
    \end{tabular}%
}  \label{tab:crimecontrC1}%
\end{table}\spacingset{\bsln}%

% Table generated by Excel2LaTeX from sheet 'Only Simpca'
\spacingset{1}
\begin{table}[h]
  \centering
  \caption{Crime data. Contributions to the second SIMPCA components. In parenthesis are shown the squared coefficients of multiple correlation.}
 {\scriptsize
    \begin{tabular}{lrr}
    \toprule
          & \multicolumn{2}{c}{comp 2} \\
\cmidrule{2-3}    Variable & \multicolumn{1}{l}{thresh} & \multicolumn{1}{l}{fwd. sel.} \\
    \midrule
    number of vacant households & 17\% (0.81) & -- \\
    number of people living in urban areas & 1\% (0.99) & 25\% (0.6) \\
    number of kids born to never married & 9\% (0.89) & -- \\
    number of people known to be foreign born & --    & 3\% (0.68) \\
    number of people in homeless shelters & 8\% (0.72) & -- \\
    number of people under the poverty level & 34\% (0.95) & -- \\
    \% of families (with kids) headed by two parents & --    & -43\% (0.51) \\
    \% of people foreign born & --    & 13\% (0.40) \\
    \% of housing units with less than 3 bedrooms & --    & 15\% (0.54) \\
    population for community & -31\% (0.99) & -- \\
    \bottomrule
    \end{tabular}%
}  \label{tab:crimecontrC2}%
\end{table}\spacingset{\bsln}%

\begin{figure}[h]
\centering
    \includegraphics[width = 0.66\textwidth]{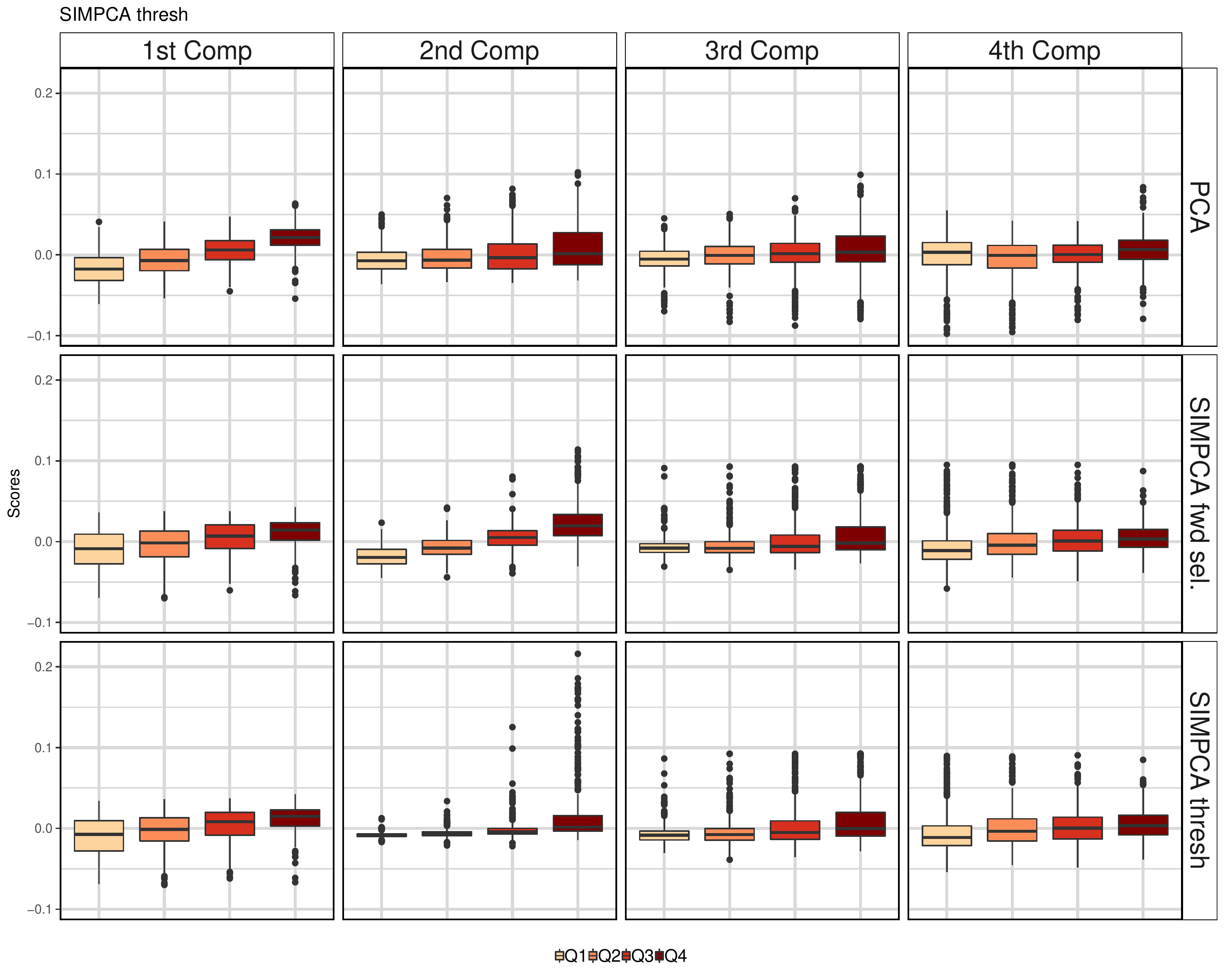}
\caption{Crime data. Distribution of the scores of the first four pcs and SIMPCA components by logarithm of rate of violent crime quartiles.}
\label{fig:crsBP}
\end{figure}

The plots in Figure \ref{fig:crsBP} show the distribution of the scores of different components with respect to the quartiles of the logarithm of violent crime. The first pc and the second SIMPCA fwd. sel. components seem to be the best at separating the quartiles. The first components computed with both SIMPCA settings also show a positive association with the response. The remaining components are rather flat with respect to the response. This impression is confirmed by the $R^2$ coefficients obtained by regressing the logarithm of violent crime on a different number of components. The values of the coefficients are quite high, considering that the regression of rate of violent crime on all one hundred variables yields a coefficient of determination equal to 0.69.

\spacingset{1}
\begin{table}[h]
  \centering
  \caption{Crime data. Coefficients of determination given by the regression of logarithm of the rate of violent crime) on up to four components computed with different methods.}
{ \scriptsize
    \begin{tabular}{lrrrr}
    \toprule
       \multicolumn{1}{l}{method} & \multicolumn{1}{l}{1 comp} & \multicolumn{1}{l}{2 comp.s} & \multicolumn{1}{l}{3 comp.s} & \multicolumn{1}{l}{4 comp.s} \\
        \midrule
        PCA   & 0.44  & 0.50  & 0.53  & 0.55 \\
        SIMPCA fwd. sel. & 0.16  & 0.53  & 0.54  & 0.54 \\
        SIMPCA thresh. & 0.16  & 0.27  & 0.33  & 0.33 \\
   \bottomrule
   \end{tabular}%
}
 \label{tab:crimeR2}%
\end{table}\spacingset{\bsln}%
%\spacingset{\bsln}  %% -------> restores baselinestretch
In summary, in this example the SIMPCA components obtained with forward selection combining few key variables show a good association with violent crime. In contrast, the components obtained by thresholding are less associated with violent crime. The variables corresponding to large coefficients of the (rotated) pcs correspond to clusters of mutually highly correlated variables which shadow other important but less correlated variables.
\subsection{Baseball Hitters data}\label{sec:baseball}
This dataset gives the statistics and salaries of different hitters who played in the 1968 U.S. Major Baseball League. The original study of these data intended to assess whether players were paid according to their performance or not. Note that the research question here is different from one addressed by a predictive model. The question concerns studying the retribution of players using a measure of their quality of play, and not to explain the retribution with their playing statistics. Hence it is required to estimate a measure of quality of play, which we expect will be useful to ``explain'' the players' salaries.

\begin{figure}[h]
\centering
\includegraphics[width = 0.66\textwidth]{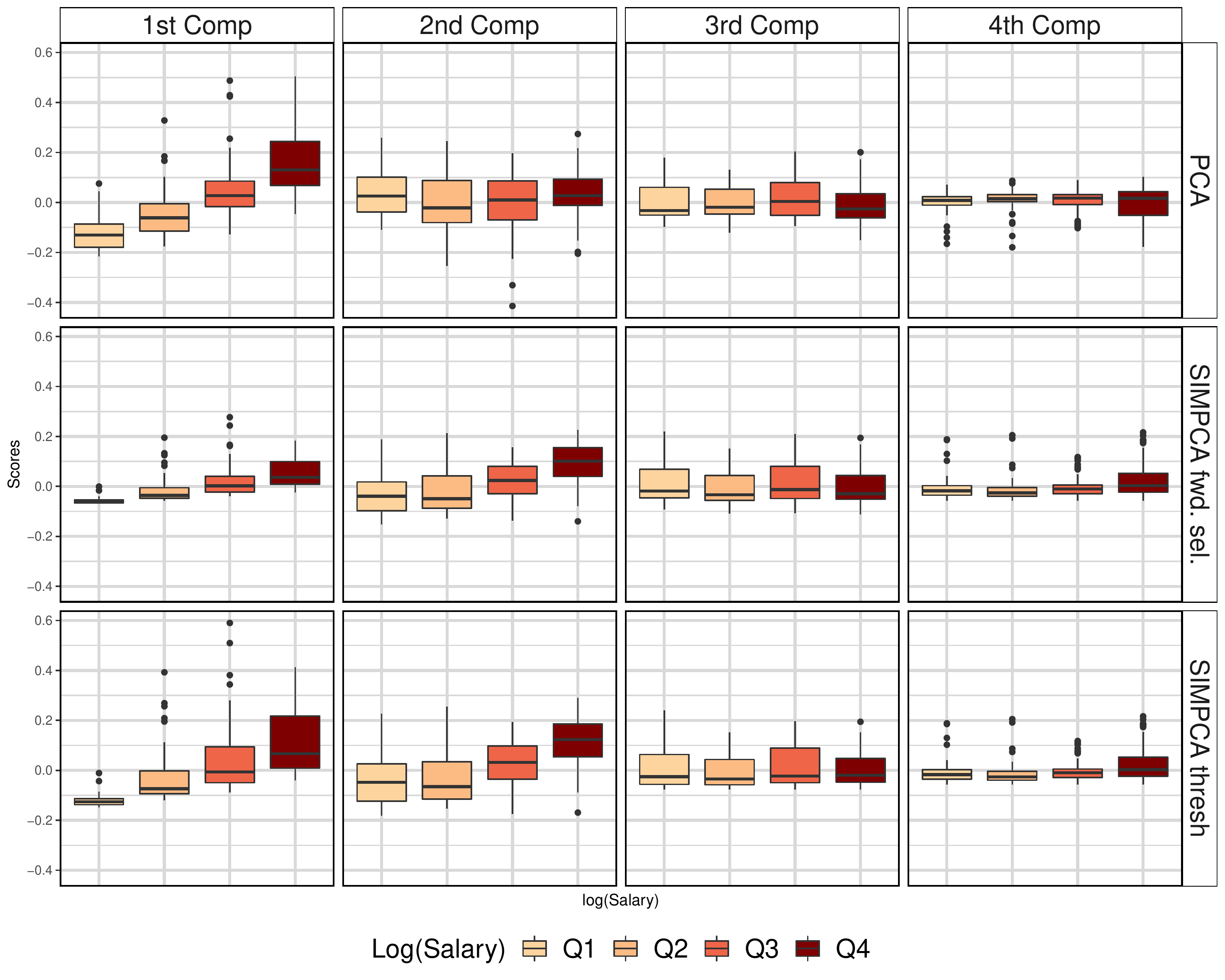}
\caption{Baseball data. Distribution of the scores of the first four pcs and SIMPCA components by log-salary quartiles of baseball hitter players.}
\label{fig:basBP}
\end{figure}

Since the statistics measured by the variables are nonhomogeneous, PCA was carried on the data scaled to unit variance. Figure \ref{fig:basBP} shows the distribution of the scores of the first four components computed with PCA and SIMPCA with respect to the quartiles of the logarithm of the salary (log(salary)). The SIMPCA components were obtained by applying varimax rotation to the pcs coefficients with Kaiser normalization. The first set of SIMPCA components (\emph{fwd sel}) was obtained using forward selection with $alpha = 99\%$ and the second set (\emph{thresh}) by thresholding the coefficients with threshold value equal to $0.3$. The log-salary quartiles are separated by the first pc but there is some overlapping, the remaining pcs are almost flat with respect to log-salary. Instead, the first two SIMPCA components, in both sets, show to achieve some separation of the salary quartiles but for both the overlapping is more pronounced than for the first pc.

The percent contributions  of the pcs and SIMPCA components showing to separate better the quartiles of the logarithm of salary in Figure \ref{fig:basBP} are shown in Table \ref{tab:bascontr}, together with the (net) variance that each of them explains and the $R^2$ coefficients of the regression of log(salaries). In the second column of the table are shown the same statistics for the projection of the first pc (PSPCA), which will be discussed later. The variables for the PSPCA component were selected to explain 99\% of the first pc's variance with forward selection.
\spacingset{1}
\begin{table}%[H]
  \centering
  \tbl{Baseball data. Percent contributions to the first pcs and different sparse components, variance explained and coefficient of determination for log(salary).}
{\scriptsize
\begin{tabular}{llllllllll}%p{1cm}p{1cm}
\toprule
      & \multicolumn{1}{c}{PCA} &       & \multicolumn{1}{c}{PSPCA} &       & \multicolumn{2}{c}{SIMPCA fwd sel} &       & \multicolumn{2}{c}{SIMPCA thresh} \\
\cmidrule{2-2}\cmidrule{4-4}\cmidrule{6-7}\cmidrule{9-10}\multicolumn{1}{c}{Variable} & \multicolumn{1}{c}{PC1} &       & \multicolumn{1}{c}{Comp 1} &       & \multicolumn{1}{c}{Comp 1} & \multicolumn{1}{c}{Comp 2} &       & \multicolumn{1}{c}{Comp 1} & \multicolumn{1}{c}{Comp 2} \\
\midrule
    \rowcolor[rgb]{ .839,  .863,  .894} times at bat in 1986 & 6\%   &       & 14\%  &       & ---   & 24\%  &       & ---   & 22\% \\
    \rowcolor[rgb]{ .839,  .863,  .894} hits in 1986 & 6\%   &       & ---   &       & ---   & ---   &       & ---   & 14\% \\
    \rowcolor[rgb]{ .839,  .863,  .894} home runs in 1986 & 6\%   &       & ---   &       & ---   & ---   &       & ---   & 15\% \\
    \rowcolor[rgb]{ .839,  .863,  .894} runs in 1986 & 6\%   &       & 11\%  &       & ---   & 28\%  &       & ---   & 17\% \\
    \rowcolor[rgb]{ .839,  .863,  .894} runs batted-in in 1986 & 7\%   &       & 10\%  &       & ---   & 36\%  &       & ---   & 18\% \\
    \rowcolor[rgb]{ .839,  .863,  .894} walks in 1986 & 6\%   &       & ---   &       & ---   & 12\%  &       & ---   & 13\% \\
    years in the major leagues & 8\%   &       & ---   &       & ---   & ---   &       & 15\%  & --- \\
    times at bat in career & 9\%   &       & ---   &       & 77\%  & ---   &       & 17\%  & --- \\
    hits in career & 9\%   &       & ---   &       & ---   & ---   &       & 14\%  & --- \\
    home runs in career & 9\%   &       & ---   &       & 23\%  & ---   &       & 11\%  & --- \\
    runs in career & 9\%   &       & ---   &       & ---   & ---   &       & 12\%  & --- \\
    runs batted-in in career & 9\%   &       & 42\%  &       & ---   & ---   &       & 15\%  & --- \\
    walks in career & 9\%   &       & 24\%  &       & ---   & ---   &       & 16\%  & --- \\
    \rowcolor[rgb]{ .839,  .863,  .894} put outs in 1986 & 2\%   &       & ---   &       & ---   & ---   &       &       & --- \\
    \rowcolor[rgb]{ .839,  .863,  .894} assists in 1986 & 1\%   &       & ---   &       & ---   & ---   &       &       & --- \\
    \rowcolor[rgb]{ .839,  .863,  .894} errors in 1986 & 0\%   &       & ---   &       & ---   & ---   &       &       & --- \\
\midrule
      &       &       &       &       &       &       &       &       &  \\
Variance explained & \multicolumn{1}{l}{46\%} &       & \multicolumn{1}{l}{46\%} &       & \multicolumn{1}{l}{42\%} & \multicolumn{1}{l}{29\%} &       & \multicolumn{1}{l}{42\%} & \multicolumn{1}{l}{29\%} \\
$R^2$ & \multicolumn{1}{l}{51\%} &       & \multicolumn{1}{l}{50\%} &       & \multicolumn{2}{c}{51\%} &       & \multicolumn{2}{c}{51\%} \\
\midrule
%\multicolumn{10}{l}{Note: A dash denotes a truly zero value.}\\
%\multicolumn{10}{l}{The gray bands separate season offensive play, career offensive play and season defensive play.} \\
\end{tabular}}
\tabnote{Note: A dash denotes a truly zero value.\hfill}
\tabnote{The gray bands separate season offensive play, career offensive play and season defensive play.\hfill}
\label{tab:bascontr}%
\end{table}\spacingset{\bsln}%

The PCA contributions are not difficult to interpret as they give almost equal weight to offensive play, with slightly higher values for career statistics, and small weight to defensive play. However, almost all statistics have nonzero coefficient.
Instead, the coefficients of the first two threholded SIMPCA components are easily interpretable. The first component is a measure of the career offensive performance with a "premium" for the years played. In this indicator the number of times at bat and the number of walks are slightly more important than the other statistics. The second component is a weighted average of the 1986 season offensive statistics.
The forward selection SIMPCA components are a simpler version of the thresholded ones and explain the same proportion of variance.
It is interesting to notice how, for both settings of SIMPCA, the career play statistics are completely separated from the season ones in the first two components and that the two sparse components together explain a much higher percentage of variance explained than the first pc, as shown in the same table.
The contributions to the forward selection SIMPCA components show that ``times at bat'' and ``home runs'' in career are enough to reproduce over 91\% of the variance explained by the first pc.

The coefficients of determination ($R^2$) for the regression of the log(salary) onto the two pairs of SIMPCA components, shown in Table \ref{tab:bascontr}, are similar to that of regressing log(salary) onto the first pc. Hence, a simple indicator of quality of play useful to explain players salary can be obtained with straight PSPCA, that is by projecting the first pc onto a set of variables which can explain 99\% of its variability. The five nonzero contributions to this component, shown in Table \ref{tab:bascontr}, illustrate that five variables are sufficient to reproduce over 99\% of the variance explained by the first pc with a negligible loss of explanatory power on the salary. In passing, we note that the variables selected with forward selection do not all correspond to ones having large coefficients in the pc.
%%% ==============  EUROJOBS ================================================
\subsection{Eurojobs data}
%%% ==============  EUROJOBS ================================================
This data set contains the percent distribution of workforce per sector in 1979 for 26 European countries. Since the measurements are all percentages we centred but did not standardise the columns. In this way the coefficients are also more easily interpretable.
The 26 countries in the data set were classified under different political blocks as follows:
%\vspace{1em}

\spacingset{1.5}
{\small
    \begin{description}
      \item[{\bf Capitalist}:]  Austria (AT), Belgium (BE), Denmark (DK), Finland (FI), France (FR), Greece (GR), Ireland (IE), Italy (IT), Luxembourg (LU), Netherlands (NL), Norway (NO),  Portugal (PT), Spain (ES), Sweden (SE), Switzerland (CH), UK (UK), West Germany (FRG);\\
     \item[{\bf Communist}:] Czechoslovakia (CZ), East Germany (DDR), Hungary (HU), Poland (PL),  Romania (RO), Soviet Union (USSR);\\
     \item[{\bf Unaligned}:] Turkey (TR), Yugoslavia (YU).
    \end{description}
}
\spacingset{\bsln}

The aim of the analysis is to find two components which can separate the countries in the three blocks.

\begin{figure}[h]
\centering
\includegraphics[width = 0.66\textwidth, keepaspectratio=True]{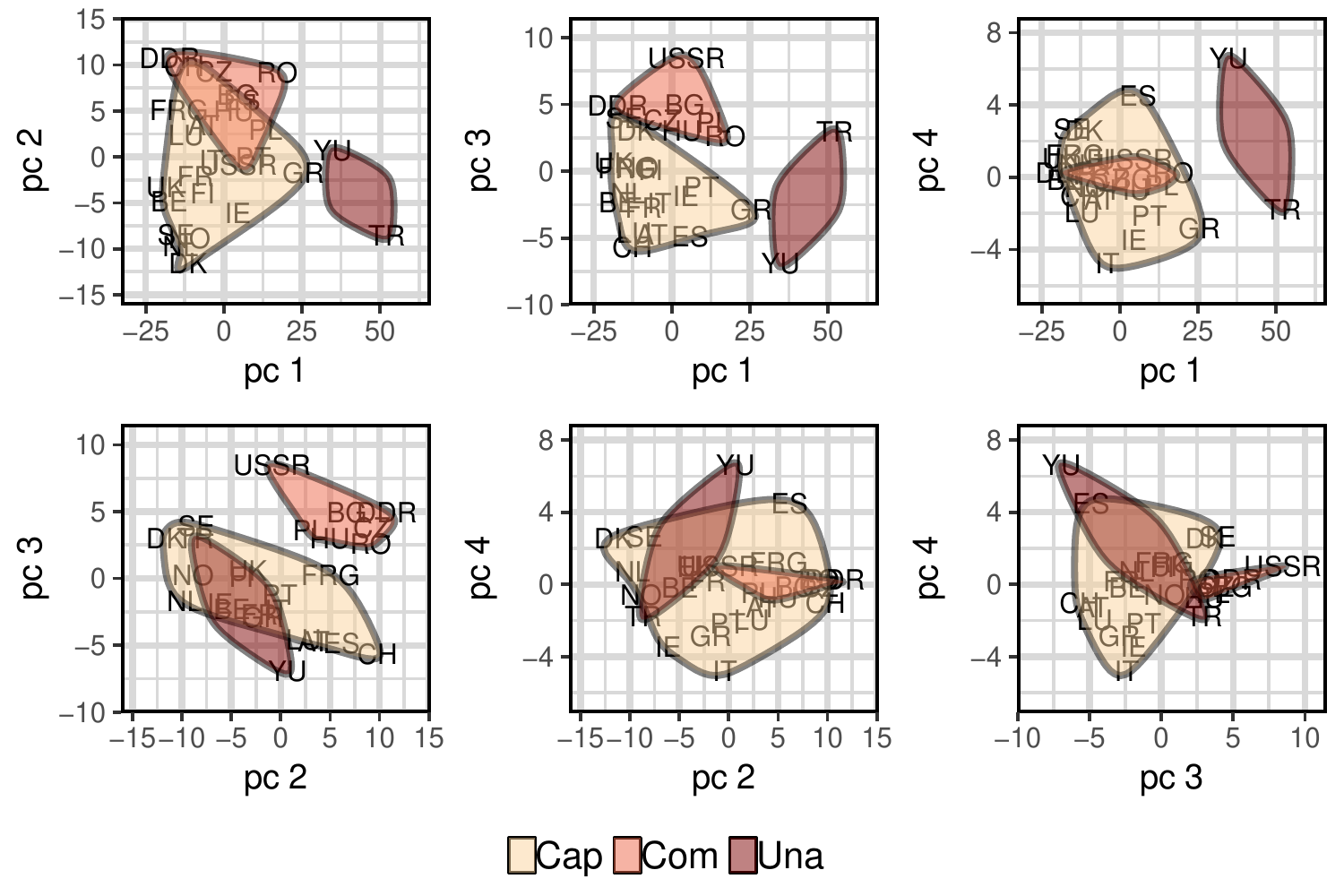}
\caption{European Job data. European countries classified by pairs of first four pcs of the Eurojobs dataset.}
\label{fig:EJscorePCs}
\end{figure}

\begin{figure}[h]
\centering
\includegraphics[width = 0.66\textwidth, keepaspectratio=True]{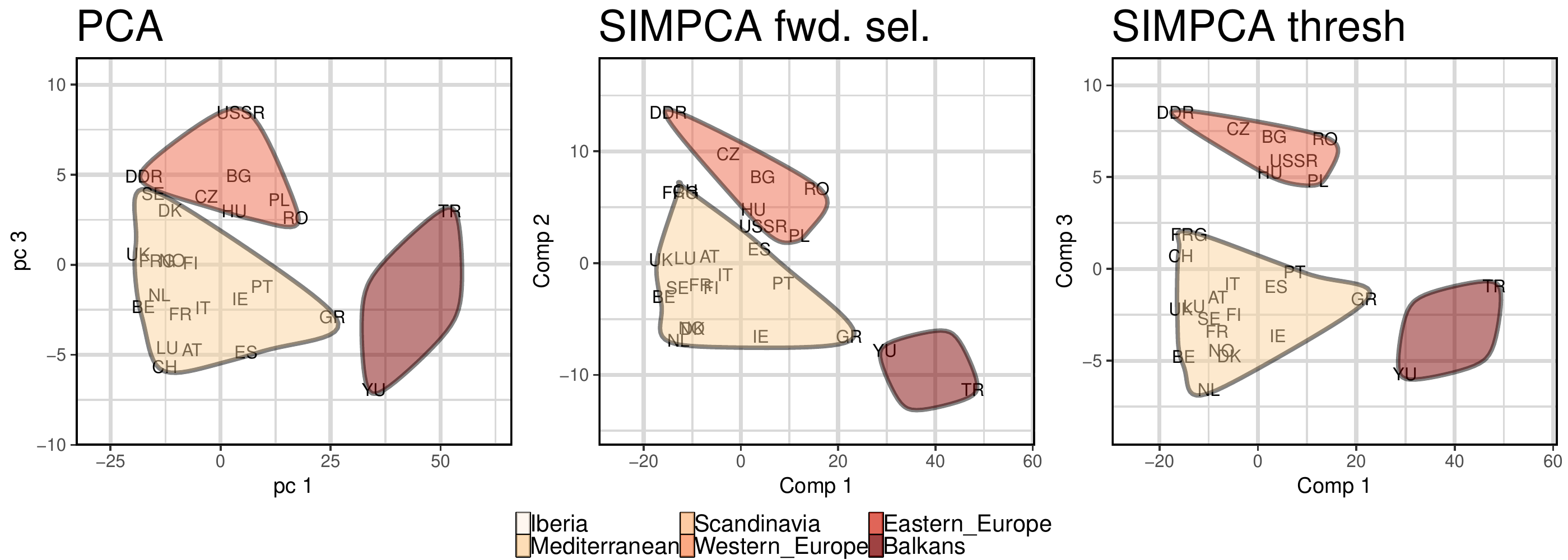}
\caption{European Job dat. European countries classified by most useful pairs of  pcs and SIMPCA components.}
\label{fig:EJsimpcascores}
\end{figure}

Figure \ref{fig:EJscorePCs} shows the paired scatter plots of the scores of the first four pcs. The countries in the Unaligned block are well separated from the others by the first pc; most of the countries in the Communist block are separated from the others by the third pc, with some overlapping (see Figure \ref{fig:EJsimpcascores} for an enlarged picture).%

%%
More interpretable and discriminatory components are those computed with two SIMPCA obtained by rotating the coefficients of the normalised pcs (that is, the coefficients divided by the corresponding singular value) using Varimax with Kaiser normalisation. In one case, denoted as \emph{SIMPCA fwd. sel}, the variables were selected with forward selection (setting $\alpha = 0.99$). In the other case, denoted as \emph{SIMPCA thresh}, the variables were selected by thresholding the  coefficients with threshold equal to 0.3.

The score plots of the pairs of pcs and SIMPCA components that best separate the blocks\footnote{The encircling polygons were computed using the function \emph{geom\_encircle} in the \texttt{R}-package \emph{ggalt}, written by Ben Bolker. The function builds an \emph{x-spline} \cite{bla} around the convex hull of the observations belonging to a given class.} are shown in Figure \ref{fig:EJsimpcascores}. The first two \emph{SIMPCA fwd. sel.} components give a pattern similar to that of the pcs but with a better separation of the unaligned countries;  the first \emph{SIMPCA thresh} component separates well the unaligned countries and the third one gives a much better separation of the Communist block from the others than any other component. Hence these last two components together achieve a complete separation of the blocks.

%%%%%%%%%%%%%%%%%%%%%%%%%%%%%%%%%%%%
Table \ref{tab:EJcontribs} shows the percent contributions to the selected pcs and SIMPCA components.
Clearly the SIMPCA components are easier to interpret: the first \emph{SIMPCA fwd. sel.} component is simply the percentage of workforce employed in agriculture; the second is the difference of the percentage employed in manufacturing and that employed in services or finance. The first \emph{SIMPCA thresh.} component is roughly the difference between the percentage of workforce employed in agriculture and the percentage of workers employed in manufacturing or services; the third component is the percentages of workers employed in services or finance decreased by about one fifth of the percentage of workers employed in manufacturing.
\spacingset{1}
\begin{table}[h]
  \centering
  \caption{European Job data. Percent contributions to two components obtained with PCA and two settings of SIMPCA that best classify countries in political blocks.}
   {\scriptsize
% Table generated by Excel2LaTeX from sheet 'Eurojobs'
\begin{tabular}{lrrrrrrrr}
\toprule
      & \multicolumn{2}{c}{PCA} &       & \multicolumn{2}{c}{SIMPCA fwd sel} &       & \multicolumn{2}{c}{SIMPCA thresh} \\
\cmidrule{1-6}\cmidrule{8-9}Sector & \multicolumn{1}{c}{pc 1} & \multicolumn{1}{l}{pc 3} &       & \multicolumn{1}{c}{Comp 1} & \multicolumn{1}{c}{Comp 2} &       & \multicolumn{1}{c}{Comp 1} & \multicolumn{1}{c}{Comp 3} \\
\cmidrule{2-3}\cmidrule{5-6}\cmidrule{8-9}agriculture & 50\%  & 5\%   &       & 100\% & --    &       & 44\%  & -- \\
mining & 0\%   & 3\%   &       & --    & --    &       & --    & -- \\
manufacturing & -15\% & 8\%   &       & --    & 51\%  &       & -27\% & 23\% \\
power supply industries & 0\%   & 0\%   &       & --    & --    &       & --    & -- \\
construction & -3\%  & -3\%  &       & --    & --    &       & --    & -- \\
service industries & -11\% & -25\% &       & --    & -28\% &       & -29\% & -33\% \\
finance & -2\%  & -21\% &       & --    & -20\% &       & --    & -44\% \\
social and personal services & -17\% & 26\%  &       & --    & --    &       & --    & -- \\
transport and communications & -3\%  & 7\%   &       & --    & --    &       & --    & -- \\
\midrule
      &       &       &       &       &       &       &       &  \\
vexp  & 82\%  & 4\%   &       & 81\%  & 10\%  &       & 80\%  & 7\% \\
\bottomrule
\end{tabular}%
}
\tabnote{Note: A dash denotes a truly zero value.\hfill}
\label{tab:EJcontribs}%
\end{table}\spacingset{\bsln}%
Since the first SIMPCA \emph{fwd sel} component (that, is the percentage of agriculture) has correlation equal to 0.99 with the first SIMPCA \emph{thresh} component, it seems reasonable to choose the former and the third SIMPCA \emph{thresh} component as the best pair to discriminate the blocks. The correlation between these two components is equal to 0.05.
\section{Conclusions}\label{sec:remarks}
In many situations interpretability of the pcs coefficients is crucial for drawing conclusions from PCA. In some cases the optimal pcs do not lend themselves to interpretation. Instead, rotated components can be more useful from this point of view. Rotating the pcs destroys their optimality but, at least at an exploratory stage of the analysis, a meaningful component can be more useful than a meaningless one which explains more variance.
The use of PCA to capture meaningful relationships of the variables with a variable external to the dataset is somewhat improper. However, not all methods used for prediction or classification produce indications on which variables are most useful for this purpose. Therefore, this approach can give useful insights and directions for further analyses.

The traditional method of ignoring "small" coefficients does not give reliable results. Instead, truly sparse components show the results after some of the variables have actually been excluded from the component, hence are more reliable. Furthermore, using a regression based variable selection approach removes the drawback of thresholding of selecting highly correlated variables. We showed in the examples how this is possible in practice.

Rotations are heuristically defined and there is no guarantee that one will give better results than another. Therefore, often several rotated results are analysed before choosing the one that best answers the research questions. SIMPCA automatises this labour intensive process and researchers can easily compute several sparse rotated solutions and compare them. The algorithm is extremely simple and can also be run as a sequence of simple operations available in most statistical software packages.

In reviewing rotation criteria we were able to prove that any orthogonal rotation applied to the unit variance coefficients (as commonly done) gives the same result as Varimax. We hope that the brief overview will be useful to practitioners.

We showed that SIMPCA yields simplified coefficients which can offer alternative characterisation of the dataset. In some cases, SIMPCA components were also better at classifying external variables than the pcs. The examples that we presented are necessarily not exhaustive of all possible applications and, in some cases, SIMPCA may not give the hoped results. However, they show, as we know from other applications we carried out, that SIMPCA has good potential for finding interesting patterns in data that cannot be found with standard PCA.
%\section*{Bibliography}
\bibliographystyle{tfs}
%\bibliography{SIMPCA_02_Master-minimal}

\begin{thebibliography}{10}
\providecommand{\MR}{\relax\unskip\space MR }
\providecommand{\url}[1]{\normalfont{#1}}
\providecommand{\urlprefix}{Available at }

\bibitem{bac}
B. Andrew, \emph{Matrix Groups An Introduction To Lie Groups}, Springer, 2002.

\bibitem{bla}
C. Blanc and C. Schlick, \emph{X-splines : A Spline Model Designed for the End
  User}, in \emph{Proceedings of SIGGRAPH 95}. ACM, 1995, p. 377–386.

\bibitem{bro}
M. Browne, \emph{An overview of analytic rotation in exploratory factor
  analysis.}, Multivariate Behavioral Research 36 (2001), pp. 111--150.

\bibitem{cad}
J. Cadima and I.T. Jolliffe, \emph{Loadings and correlations in the
  interpretation of principal components}, Journal of Applied Statistics 22
  (1995), pp. 203--214.

\bibitem{cra}
C.B. Crawford and G.A. Ferguson, \emph{A general rotation criterion and its use
  in orthogonal rotation.}, Psychometrika 35 (1970), pp. 321--332.

\bibitem{die}
J. Dien, \emph{Evaluating two-step pca of erp data with geomin, infomax,
  oblimin, promax, and varimax rotations}, Psychophysiology 47 (2010), pp.
  170--183.

\bibitem{uci}
A. Frank and A. Asuncion, \emph{{UCI} machine learning repository}, University
  of California, Irvine, School of Information and Computer Sciences (2010).
  \url{http://archive.ics.uci.edu/ml}.

\bibitem{har}
H. Harman, \emph{{Modern Factor Analysis}}, University Of Chicago Press, 1976.

\bibitem{has}
T. Hastie, R. Tibshirani, and J. Friedman, \emph{The elements of statistical
  learning: data mining, inference and prediction}, 2nd ed., Springer, 2009.

\bibitem{hot}
H. Hotelling, \emph{{Analysis of a Complex of Statistical Variables with
  Principal Components}}, Journal of Educational Psychology 24 (1933), pp.
  498--520.

\bibitem{ize}
A.J. Izenman, \emph{{Modern Multivariate Statistical Techniques : Regression,
  Classification, and Manifold Learning}}, Springer Texts in Statistics,
  Springer New York, 2008.

\bibitem{jac}
J.E. Jackson, \emph{A user's guide to principal components}, Vol. 587, John
  Wiley \& Sons, 2005.

\bibitem{jaco}
N. Jacobs and D. Harvey, \emph{Do parents make a difference to children’s
  academic achievement? differences between parents of higher and lower
  achieving students}, Educational studies 31 (2005), pp. 431--448.

\bibitem{jef}
J. Jeffers, \emph{Two case studies in the application of principal component.},
  Applied Statistics 16 (1967), pp. 225--236.

\bibitem{jen}
R. Jennrich, \emph{A simple general procedure for orthogonal rotation.},
  Psychometrika 66 (2001), pp. 289--306.

\bibitem{jen02}
R. Jennrich, \emph{A simple general method for oblique rotation}, Psychometrika
  67 (2002), pp. 7--19.

\bibitem{jol95}
I. Jolliffe, \emph{Rotation of principal components: Choice of normalization
  constraints}, Journal of Applied Statistics 22 (1995), pp. 29--35.

\bibitem{jol}
I. Jolliffe, \emph{Principal component analysis}, 2nd ed., Springer series in
  statistics, Springer-Verlag, 2002.

\bibitem{jol03}
I. Jolliffe, N. Trendafilov, and M. Uddin, \emph{A modified principal component
  technique based on the lasso}, Journal of Computational and Graphical
  Statistics 12 (2003), pp. pp. 531--547.

\bibitem{jol00}
I. Jolliffe and M. Uddin, \emph{The simplified component technique: An
  alternative to rotated principal components}, Journal of Computational and
  Graphical Statistics 9 (2000), pp. 689--710.

\bibitem{jou}
M. Journ\'{e}e, Y. Nesterov, P. Richt\'{a}rik, and R. Sepulchre,
  \emph{Generalized power method for sparse principal component analysis.},
  Journal of Machine Learning Research 11 (2010), pp. 517--553.

\bibitem{kai}
H. Kaiser, \emph{The varimax criterion for analytic rotation in factor
  analysis.}, Psychometrika 23 (1958), pp. 187--200.

\bibitem{hen}
H.A.L. Kiers, \emph{Three-mode orthomax rotation}, Psychometrika 62 (1997), pp.
  579--598.
  \urlprefix\url{http://gen.lib.rus.ec/scimag/index.php?s=10.1007/bf02294644}.

\bibitem{kie}
H.A. Kiers and I.V. Mechelen, \emph{Three-way component analysis: Principles
  and illustrative application.}, Psychological methods 6 (2001), p.~84.

\bibitem{kou}
T. Kourti and J.F. MacGregor, \emph{Process analysis, monitoring and diagnosis,
  using multivariate projection methods}, Chemometrics and intelligent
  laboratory systems 28 (1995), pp. 3--21.

\bibitem{mer15}
G. Merola, \emph{Least squares sparse principal component analysis: a backward
  elimination approach to attain large loadings}, Australia \& New Zealand
  Journal of Statistics 57 (2015), pp. 391--429.

\bibitem{mer19}
G.M. Merola and G. Chen, \emph{Projection sparse principal component analysis:
  An efficient least squares method}, Journal of Multivariate Analysis 173
  (2019), pp. 366 -- 382.
  \urlprefix\url{http://www.sciencedirect.com/science/article/pii/S0047259X18305463}.

\bibitem{mog}
B. Moghaddam, Y. Weiss, and S. Avidan, \emph{Spectral Bounds for Sparse PCA:
  Exact and Greedy Algorithms}, in \emph{Advances in Neural Information
  Processing Systems}. MIT Press, 2006, pp. 915--922.

\bibitem{pea}
K. Pearson, \emph{On lines and planes of closest fit to systems of points in
  space}, The London, Edinburgh, and Dublin Philosophical Magazine and Journal
  of Science 2 (1901), pp. 559--572.

\bibitem{ric}
M.B. Richman, \emph{Rotation of principal components}, International Journal of
  Climatology 6 (1986).
  \urlprefix\url{http://gen.lib.rus.ec/scimag/index.php?s=10.1002/joc.3370060305}.

\bibitem{sri}
B.K. Sriperumbudur, D.A. Torres, and G. Lanckriet, \emph{Sparse eigen methods
  by D.C. programming}, in \emph{Proceedings of the 24th international
  conference on Machine learning}. ACM Press, 2007, pp. 831--838.

\bibitem{thu}
L. Thurstone, \emph{Multiple-factor analysis}, The University of Chicago Press,
  1947.

\bibitem{vbu}
S. van  Buuren and K. Groothuis-Oudshoorn, \emph{mice: Multivariate imputation
  by chained equations in r}, Journal of Statistical Software 45 (2011), pp.
  1--67.

\bibitem{zou}
H. Zou, T. Hastie, and R. Tibshirani, \emph{Sparse principal component
  analysis}, Journal of Computational and Graphical Statistics 15 (2006), pp.
  265--286.

\end{thebibliography}

\appendix
\section{Data sources}\label{app:data}
The data used in the examples were taken from the StatLib data archives (\url{http://lib.stat.cmu.edu/datasets/})
and its "Data And Story Library" (DASL, \url{http://lib.stat.cmu.edu/DASL/DataArchive.html}) or from the University of California, Irvine, Machine Learning Repository (UCI, \url{http://archive.ics.uci.edu/ml/}) \citep{uci}.
\subsection*{\bf Human Freedom}
The set was used to build the 2016 Human Freedom index and contains 79 distinct indicators of personal and economic freedom in different following areas. The indicators measure freedom on a scale from zero to ten, where ten represents more freedom. The dataset is available at\\
\url{https://www.cato.org/human-freedom-index-new}

Analysis of the data is available in\\
Vasquez, I. and Porcnik, T. 2016. The Human freedom Index 2016. Cato Institute, the Fraser Institute, and the Friedrich Naumann Foundation for Freedom.

\subsection*{\bf Crime data}
The set contains 122  socio-economic indicators  and the corresponding per capita rate of violent crime for 1994 US cities. The data were taken in the early 1990's. They were donated by Michael Redmond (redmond@lasalle.edu); Computer Science; La Salle University; Philadelphia, PA, 19141, USA.

The data were downloaded from UCI:\\
\url{http://archive.ics.uci.edu/ml/machine-learning-databases/communities/}.\\
Details can be fond in \\
\url{http://archive.ics.uci.edu/ml/machine-learning-databases/communities/communities.names/}.\\

There is no published analysis of these data. The data sources are given below.

\noindent
References\\
U. S. Department of Commerce, Bureau of the Census, Census Of Population And Housing
1990 United States: Summary Tape File 1a \& 3a (Computer Files)

\noindent
U.S. Department Of Commerce, Bureau Of The Census Producer, Washington, DC and
Inter-university Consortium for Political and Social Research Ann Arbor, Michigan.
(1992)

\noindent
U.S. Department of Justice, Bureau of Justice Statistics, Law Enforcement Management
And Administrative Statistics (Computer File) U.S. Department Of Commerce, Bureau Of
The Census Producer, Washington, DC and Inter-university Consortium for Political and
Social Research Ann Arbor, Michigan. (1992)

\noindent
U.S. Department of Justice, Federal Bureau of Investigation, Crime in the United
States (Computer File) (1995)

\noindent
Redmond, M. A. and A. Baveja: A Data-Driven Software Tool for Enabling Cooperative
Information Sharing Among Police Departments. European Journal of Operational Research
141 (2002) 660-678.
\subsubsection*{\bf Baseball data}
The dataset contains performance statistics and salary of North American Major League Baseball players during the 1986 season. It was presented in 1988 at the ASA Graphics Section Poster Session, orgainised by Lorraine Denby. We used the original file, "baseball.data" applying the corrections suggested in [a], for the hitters only.

These data are available from the StatLib Data Archive:\\
\url{http://lib.stat.cmu.edu/datasets/baseball.data}

\noindent
Reference\\%

\noindent
[a] D.C. Hoaglin and P.F. Velleman, \emph{A critical look at some analyses of major league baseball
salaries}, The American Statistician 49 (1995), pp. 277–-285.

\subsection*{\bf Eurojobs data}
The data are the percentage employed in different industries in Europe countries during 1979. Multivariate techniques, such as cluster analysis and principal component analysis, may be used to examine which countries have similar employment patterns.

\noindent
References\\
Euromonitor (1979), European Marketing Data and Statistics, London: Euromonitor Publications, 76-77. Also found in: Manly, B.F.J. (1986) Multivariate Statistical Methods: A Primer, New York: Chapman \& Hall, 11. Also found in: Hand, D.J., et al. (1994) A Handbook of Small Data Sets, London: Chapman \& Hall, 303.

This data set is available at DASL:\\
Data \url{http://lib.stat.cmu.edu/DASL/Datafiles/EuropeanJobs.html}\\
Story \url{http://lib.stat.cmu.edu/DASL/Stories/EuropeanJobs.html}
\subsubsection*{\bf Protein Data}
The data represent estimates of the protein consumption per inhabitants for 25 European countries (see Greenacre, 1984, table 9.10); the data were originally reported by Weber, 1973, in a mimeograph published at Kiel University, Institut für Agrarpolitik und Marktlehre, entitled "Agrarpolitik im Spannungsfeld der Internationalen Ernährungspolitik". Thus, the data are not frequencies, but they are analogous to frequencies in that a total mass of protein is distributed over the cells of the matrix in units of 0.1 gram (per head per day).

This data set is available at DASL:\\
Data \url{http://lib.stat.cmu.edu/DASL/Datafiles/Protein.html}\\
Story \url{http://lib.stat.cmu.edu/DASL/Stories/ProteinConsumptioninEurope.html}.

\noindent
Reference\\
Greenacre, M.J. (1984). Theory and Applications of Correspondence
Analysis. London: Academic Press.
\end{document}